\documentclass[Afour,sageh,times]{sagej}

\usepackage{amsmath}
\usepackage{graphicx}
\graphicspath{{./}}
\usepackage{tikz}
\usepackage{tikz-3dplot}
\usetikzlibrary{shapes.geometric}
\usetikzlibrary{calc}

\definecolor{markC0}{HTML}{1F77B4}
\definecolor{markC1}{HTML}{FF7F0E}
\definecolor{markC2}{HTML}{2CA02C}
\definecolor{markC4}{HTML}{9467BD}
\definecolor{tblviolet}{HTML}{54278F}
\newcommand{\markCircle}{\,\tikz[baseline=-0.55ex]{\node[circle,fill=markC0,draw=black,line width=0.25pt,inner sep=1.6pt]{};}}
\newcommand{\markSquare}{\,\tikz[baseline=-0.55ex]{\node[rectangle,fill=markC1,draw=black,line width=0.25pt,minimum size=5pt,inner sep=0pt]{};}}
\newcommand{\markDiamond}{\,\tikz[baseline=-0.55ex]{\node[diamond,fill=markC2,draw=black,line width=0.25pt,minimum width=4.5pt,minimum height=6.5pt,inner sep=0pt]{};}}
\newcommand{\markStar}{\,\tikz[baseline=-0.55ex]{\node[star,star points=5,fill=markC4,draw=black,line width=0.25pt,minimum size=7pt,inner sep=0pt]{};}}
\usepackage{bm}
\usepackage{subcaption}
\usepackage{listings}
\usepackage{multicol}
\usepackage{enumitem}
\usepackage{booktabs}
\usepackage{colortbl}
\usepackage{xurl}   
\usepackage{float}
\usepackage{pgfplotstable}
\pgfplotstableset{col sep=comma}
\usepackage{cleveref}
\crefname{lstlisting}{Listing}{Listings}
\Crefname{lstlisting}{Listing}{Listings}
\newcommand{\projectname}{TERRA-NG}

\makeatletter
\renewcommand\subsection{\@startsection{subsection}{2}{\z@}%
  {0.9\@bls plus .3\@bls minus .1\@bls}%
  {4pt\@afterindentfalse}%
  {\sagesf\large\bfseries\raggedright}}
\makeatother

\newif\ifshowcomments
\showcommentstrue

\definecolor{fbdarkgreen}{HTML}{006400}
\ifshowcomments
  \newcommand{\nk}[1]{\textcolor{blue}{nk: #1}}
  \newcommand{\gr}[1]{\textcolor{magenta}{gr: #1}}
  \newcommand{\fb}[1]{\textcolor{fbdarkgreen}{fb: #1}}
  \newcommand{\mm}[1]{\textcolor{red}{mm: #1}}
\else
  \newcommand{\nk}[1]{}
  \newcommand{\gr}[1]{}
  \newcommand{\fb}[1]{}
  \newcommand{\mm}[1]{}
\fi

\begin{document}

\runninghead{B\"ohm et al.}

\title{Performance Analysis of Low-Order, GPU-accelerated Finite Element Kernels using Kokkos}

\author{Fabian B\"ohm\affilnum{1},
        Nils Kohl\affilnum{2},
        Harald K\"ostler\affilnum{1,3},
        Ulrich R\"ude\affilnum{3,4,5}\\[0.6em]
        \normalfont\small
        \affilnum{1}Erlangen National High Performance Computing Center (NHR@FAU), Erlangen, Germany\\
        \affilnum{2}Department of Earth and Environmental Sciences, LMU Munich, Germany\\
        \affilnum{3}Chair for System Simulation (CS10), Friedrich--Alexander--Universit\"at Erlangen--N\"urnberg, Erlangen, Germany\\
        \affilnum{4}CERFACS, Toulouse, France\\
        \affilnum{5}V\v{S}B Technical University of Ostrava, Ostrava, Czech Republic}

\affiliation{$^{1}$Erlangen National High Performance Computing Center (NHR@FAU), Erlangen, Germany\\
$^{2}$Department of Earth and Environmental Sciences, LMU Munich, Germany\\
$^{3}$Chair for System Simulation (CS10), Friedrich--Alexander--Universit\"at Erlangen--N\"urnberg, Erlangen, Germany\\
$^{4}$CERFACS, Toulouse, France\\
$^{5}$V\v{S}B Technical University of Ostrava, Ostrava, Czech Republic}

\corrauth{Fabian B\"ohm, Erlangen National High Performance Computing Center (NHR@FAU), Erlangen, Germany.}

\email{fabian.boehm@fau.de}

\begin{abstract}
We study performance portability for low-order, matrix-free finite element kernels, using the example of a vectorial, variable-coefficient PDE operator originating in geophysical models.
Written in Kokkos, the kernel is compared on NVIDIA H100, AMD MI250X, AMD MI300A and Intel PVC Max 1550 GPUs.
Owing to its low order and to optimizations that reduce the arithmetic, the kernel has a low arithmetic intensity, so that its performance is determined by how the finite element assembly is mapped onto the memory hierarchy. This is a dimension in which the architectures differ even within one vendor family, causing different performance characteristics.
We examine how Kokkos' hierarchical parallelism and shared scratch memory, which are used for the shared degrees of freedom of the conforming discretization, behave on each device. Finally, we show how portability gaps can be narrowed with tuning levers such as the size of the thread groups, the balance between occupancy and register use, and the atomic accumulation strategy at the end of the kernel.
\end{abstract}


\maketitle

\section{Introduction}

Numerical simulation at scale requires powerful supercomputers, and the supercomputer hardware landscape has 
turned toward GPU-based systems. 
The European supercomputers span all three GPU vendors: NVIDIA in JUPITER
(J\"ulich, Germany) and MareNostrum~5 (Barcelona, Spain), AMD in LUMI-G
(Kajaani, Finland) and Hunter (Stuttgart, Germany), and Intel in SuperMUC-NG
Phase~2 (Munich, Germany). An
application written in architecture-specific code is confined to the machines of one vendor. 
Exploiting all available resources in contrast requires portability across vendors. A single portable source, however, does not
imply a single performance profile: GPU architectures differ in arithmetic throughput, cache
hierarchy, coherence and atomic throughput, so the same kernel can 
achieve maximal {\em roofline} performance on one card and still be far below it on another.

Several studies have compared finite-element GPU kernels across
architectures. The CEED benchmarks compare matrix-free operator
evaluation across codes and
platforms~\citep{fischer2020scalability,kolev2021efficient}. hipBone
provides a portable Nek5000 proxy benchmarked on NVIDIA and AMD
GPUs~\citep{chalmers2023hipbone}, the MARBL hydrodynamics code was converted from
assembled matrices to MFEM's matrix-free partial assembly, with RAJA kernel
abstractions and Umpire memory management~\citep{vargas2022mfem}, Albany demonstrated
Kokkos-based~\citep{trott2022kokkos,edwards2014kokkos} portability of
finite-element assembly across NVIDIA GPUs,
Xeon Phi, and CPUs~\citep{demeshko2019albany}, deal.II provides a generic
matrix-free framework~\citep{Kronbichler:2012:CAF} that was extended to
GPUs~\citep{kronbichler2019,arndt2025dealii}, and
\citet{settgast2023lowordergpu} presented matrix-free Laplace and
elasticity kernels on V100, A100, and MI250X for geoscientific
applications. HOMMEXX rewrote the spectral-element atmosphere
dynamical core of E3SM in C++/Kokkos~\citep{bertagna2019hommexx}. Kokkos codes have also been compared across
the GPUs outside the finite-element setting:
LAMMPS-KOKKOS reports molecular dynamics on H100, MI250X, MI300A and
PVC~\citep{johansson2025lammps}, and kokkos-fft reports a
spectral turbulence solver on A100, H100, MI250X and
PVC~\citep{asahi2025kokkosfft}.

The kernels studied in this paper originate from global mantle convection models.
The Earth's mantle is a ${\sim}2900$\,km thick layer of rock between crust
and core that, despite its high viscosity, flows over geologic time. This
creeping motion drives plate tectonics and shapes the planet's thermal
evolution~\citep{schubert2001}. 
The flow is governed by a variant of Stokes (zero-Reynolds-number)
equation combined with energy transport at very high Rayleigh number: 
the circulation is 
characterized by thin
thermal boundary layers, narrow plumes, and sinking
slabs~\citep{schubert2001}, structures that become very small (relative to domain size) with high Rayleigh number. 
Resolving them requires a grid spacing at the kilometre 
scale over the whole mantle.
At that resolution the discretised systems reach on the order of $10^{12}$
unknowns per timestep~\citep{bauer2019,boehm2026terrang}. Assembling a sparse matrix at
this size exceeds the main memory of the
largest existing machines. 
Therefore, matrix-free operator evaluation~\citep{may2015,Kronbichler:2012:CAF}, which recomputes element contributions on the fly, is mandatory. 
Moreover, the mantle's viscosity
varies by several orders of magnitude over a few hundred
kilometres~\citep{lin2020viscosity,stotz2017viscosity}, which can be numerically 
challenging~\citep{Deubelbeiss:2008:PEPI,Heister:2017:GJI}. 
Such steep, near-discontinuous coefficients undercut the accuracy advantage of high-order
elements, making low-order discretisations the method of choice. 

\subsection*{Contribution}
We see the following gap in the literature highlighted above: the cross-code comparisons are dominated by high-order operators with high arithmetic intensity. 
On the other hand, 
existing studies that 
analyze the performance of low-order matrix-free kernels on GPUs 
are with few exceptions limited to simple, constant-coefficient operators. 

This work contributes a study of performance portability through Kokkos~\citep{trott2022kokkos,edwards2014kokkos} for a low-order, matrix-free Finite Element kernel, compared across the GPU
architectures of all three major vendors: NVIDIA H100, AMD MI250X and
MI300A, and Intel Ponte Vecchio.
We consider a production setting, where the kernel implements complex physics, that is a vectorial, variable-coefficient viscous operator and is highly optimized, e.g. the kernel uses the hierarchical
team parallelism of Kokkos, and stages shared DoFs from the conforming discretization in team scratch memory.
These are features on which portability depends, since the
resources they map onto, i.e.~the memory hierarchy of the card, differ substantially between architectures. 

\Cref{sec:grid} introduces the discretisation and the kernel.
\Cref{sec:opts} presents the domain-specific optimisations that take it from
a textbook baseline to the production kernel, 
detailing and justifying our claim of a high level of optimization, 
and \Cref{sec:cross-vendor} compares the kernel across the four architectures in
terms of roofline placement, throughput, and memory utilisation.

\section{TERRA-NG in a nutshell}
\label{sec:grid}
The kernel's structure follows from the discretisation and the solver it serves, so we summarise both here. 
A companion paper~\citep{boehm2026terrang} describes \projectname{},
i.e.~the geophysics code in which the new GPU-kernels are eventually employed,
including a more detailed description of the application-specific discretization with wedges,
the solver hierarchy, 
the physical validation 
against community benchmarks, and results from
production-level mantle-convection simulations at the scale of $10^{12}$ Degrees of freedom and 1 Kilometer between gridpoints spatial resolution. The current article here, in contrast, is focussed on the extensive performance and portabilty analysis of the central GPU-compute kernels, including detailed techniques of code optimization.
Every kernel discussed
here, including the variants of \Cref{tab:opts}, is available at
\url{https://github.com/mantleconvection/TERRA-NG}.

\projectname{} solves the coupled Stokes--energy system of mantle
convection, with momentum and mass balance
\begin{equation}
\begin{aligned}
  \underbrace{-\nabla \cdot \Bigl[\, 2\eta \bigl( \boldsymbol{\varepsilon}(\mathbf{u})
  - \tfrac{1}{3}\,(\nabla \cdot \mathbf{u})\,\mathbf{I} \bigr) \Bigr]}_{\text{viscous operator}}
  + \nabla p' &= \bar\rho\,\bar\alpha\,T'\,\hat{\mathbf{r}}, \\[0.4em]
  \nabla \cdot \bigl(\bar\rho\,\mathbf{u}\bigr) &= 0,
\end{aligned}
\label{eq:stokes}
\end{equation}
for velocity $\mathbf{u}$ and pressure perturbation $p'$, with the symmetric
gradient $\boldsymbol{\varepsilon}(\mathbf{u}) = \tfrac{1}{2}\bigl(\nabla\mathbf{u}
+ (\nabla\mathbf{u})^{\!\top}\bigr)$, viscosity
$\eta(\mathbf{x})$ and reference density and expansivity profiles
$\bar\rho(r)$, $\bar\alpha(r)$.

\begin{figure*}[t]
  \centering
  \includegraphics[width=0.24\textwidth]{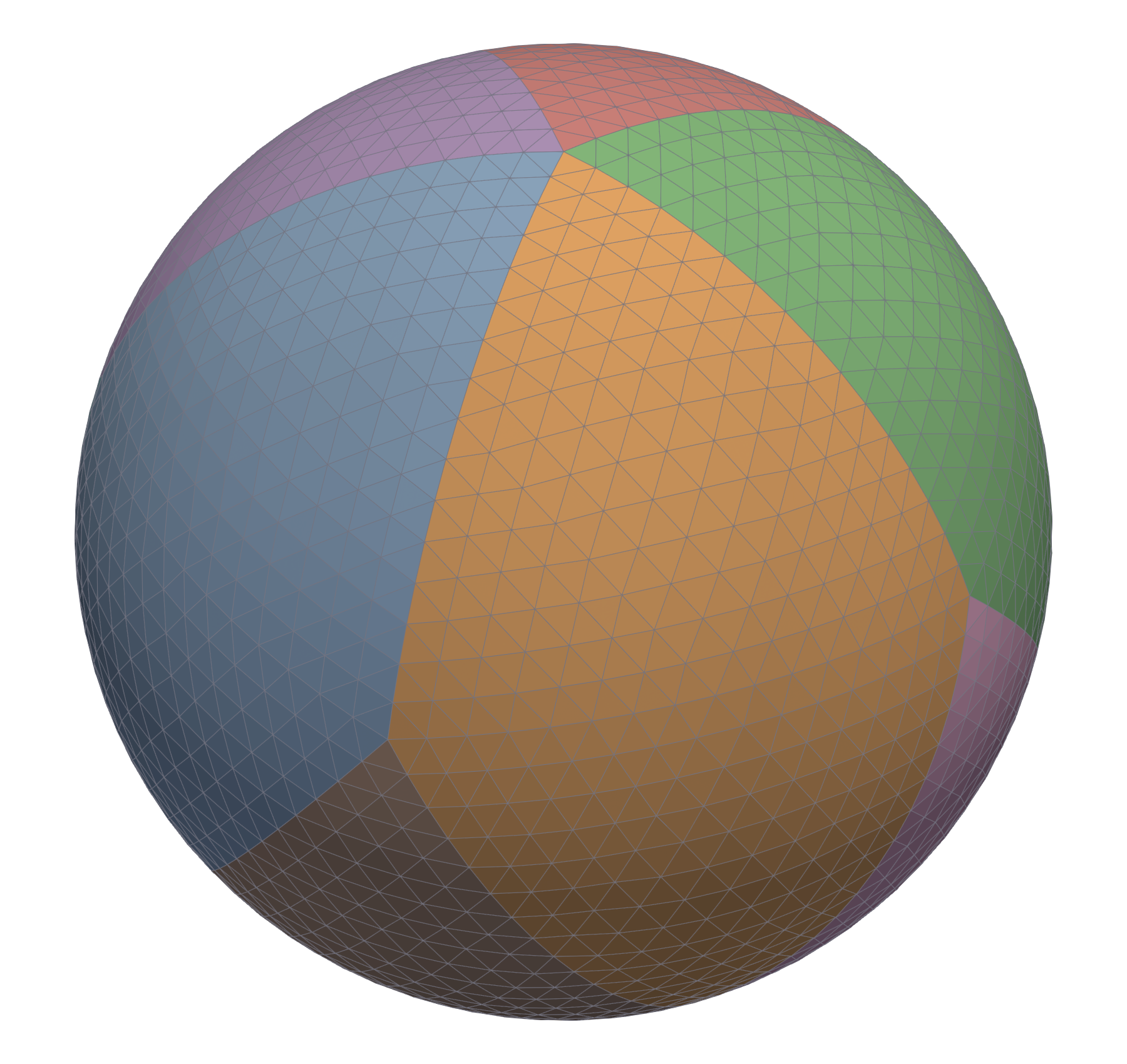}%
  \hspace{0.08\textwidth}%
  \includegraphics[width=0.34\textwidth]{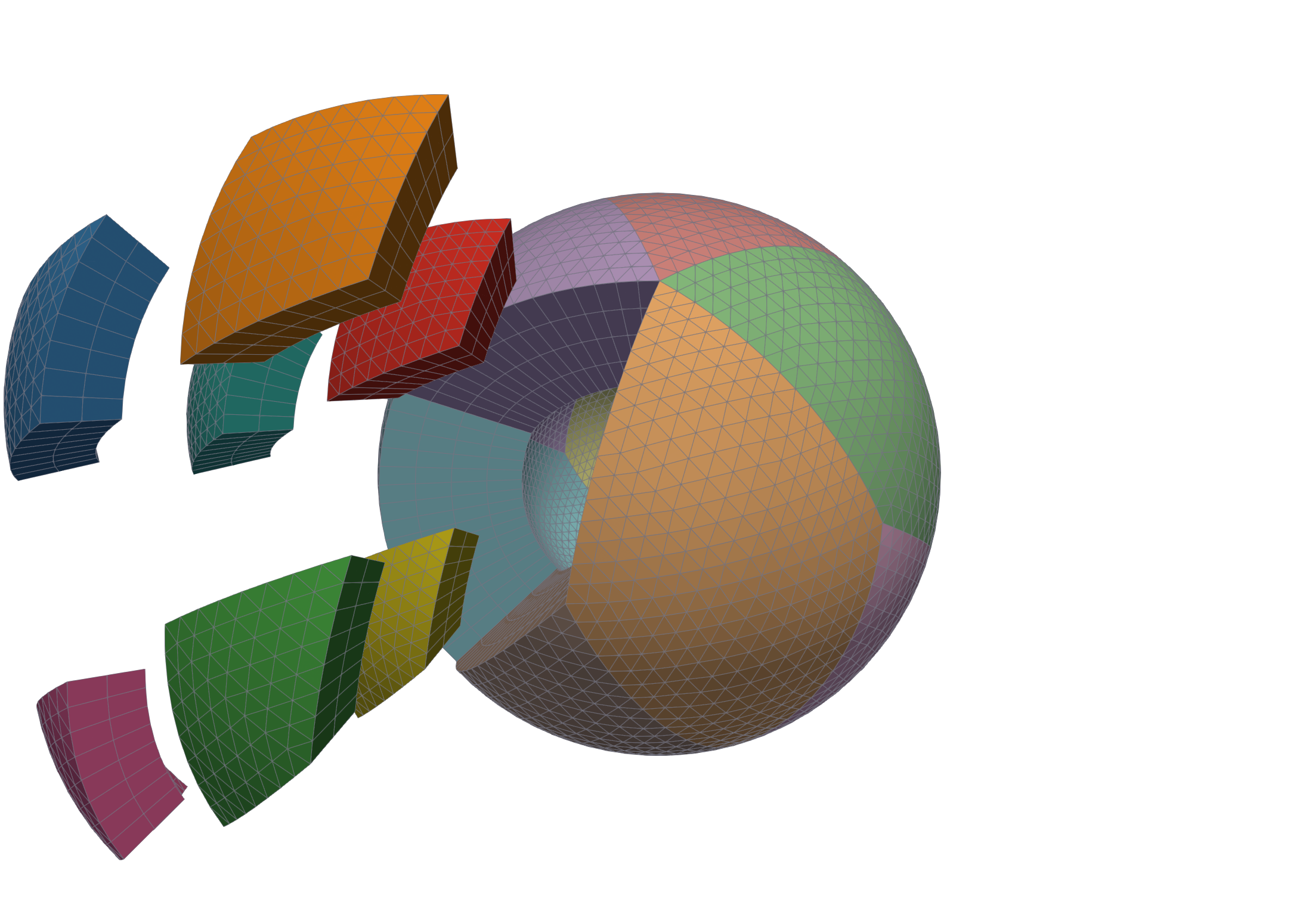}
  \caption{Left: the ten spherical diamonds of the icosahedral surface
  grid. Right: the radially extruded shell, exploded into the subdomains
  distributed across MPI ranks.}
  \label{fig:gridB}
\end{figure*}

\paragraph{Grid.}
The Earth's 
spherical surface is partitioned into ten spherical diamonds spanned by
the twelve vertices of an icosahedral base grid, 
refined by recursive bisection into $2^\ell \times 2^\ell$ curved quadrilaterals per diamond,
and extruded through $2^{\ell-1}$ radial layers. Each hexahedral cell is split implicitly into two wedge elements (\Cref{fig:gridB}). 
The mesh is  logically structured within each diamond, so an MPI rank holds the unknowns
of its subdomains in a single Kokkos 4D view, indexed by (subdomain, lateral $x$,
lateral $y$, radial layer), and recomputes the element geometry on the fly
from a 2D lateral coordinate field and a 1D radial profile. 
Within a rank, Kokkos thread teams own radially extruded tiles of a subdomain, explained in
\Cref{sec:opts}.

\paragraph{Discretisation.}
Momentum and mass balance are discretised on wedge elements, using the inf-sup stable
$W_1$-iso-$W_2/W_1$ pair, where $W_k$ denotes the wedge element space spanned
by the tensor product of a degree-$k$ triangular and a degree-$k$ radial
basis: velocity on the wedge mesh at level $\ell$, pressure one level
coarser. On the reference wedge ($\xi,\eta \ge 0$, $\xi + \eta \le 1$,
$\zeta \in [-1,1]$) the six scalar shape functions are the tensor products
\begin{equation}
  N_j \;=\; N^{\mathrm{lat}}_{j \bmod 3}(\xi,\eta)\;
         N^{\mathrm{rad}}_{\lfloor j/3 \rfloor}(\zeta), \qquad j = 0,\dots,5,
\label{eq:basisB}
\end{equation}
with $N^{\mathrm{lat}}_0 = 1{-}\xi{-}\eta$, $N^{\mathrm{lat}}_1 = \xi$,
$N^{\mathrm{lat}}_2 = \eta$ and $N^{\mathrm{rad}}_0 = \tfrac12(1{-}\zeta)$,
$N^{\mathrm{rad}}_1 = \tfrac12(1{+}\zeta)$, associating nodes $0$--$2$ with
the bottom and $3$--$5$ with the top triangular face. Each wedge holds
six nodes and three velocity components, giving eighteen velocity unknowns per
element. No global matrix is
assembled: e.g. the viscous operator in \cref{eq:stokes} is applied matrix-free with element matrices
\begin{equation}
  (A_e)_{ij} = \int_{\Omega_e} \mu \Big[ 2\,\bm{\varepsilon}(\bm{\varphi}_i) : \bm{\varepsilon}(\bm{\varphi}_j) - \tfrac{2}{3}\,\mathrm{div}(\bm{\varphi}_i)\,\mathrm{div}(\bm{\varphi}_j) \Big] \,\mathrm{d}x
  \label{eq:elemmat}
\end{equation}
evaluated on-the-fly per wedge.

The remainder of this paper focuses on the viscous operator. It is the
component applied most often in a solve: \projectname{} preconditions the
velocity block with a multigrid V-cycle whose Chebyshev smoother applies it
several times on every level of every cycle, so its cost dominates the Stokes
solve.

\paragraph{Kernel Pseudo-Code.}
Listing~\ref{lst:kernel} is the textbook variant of the matrix-free finite
element operator, the baseline for the optimizations studied in the rest of the
paper. Its members $\mathtt{src\_}$ and $\mathtt{dst\_}$ are
\texttt{Grid4DDataVec}, our alias for a set of four-dimensional
\texttt{Kokkos::Views} indexed by $(sd, x, y, r)$, the subdomain, lateral and
radial index, one per velocity component. The operator is applied as a single
\texttt{parallel\_for} on each GPU and rank, followed by a fence and one MPI
exchange of the halo layers.

The body of that dispatch, \texttt{operator()}, is the per-hex-cell work and the
subject of the rest of this paper. One thread handles one hex cell: it gathers
that hex cell's source DoFs
$\mathtt{src\_e}$, 8 nodes times 3 velocity components, so $24$
values. It then walks both wedges and their quadrature points, evaluating the
Jacobian and the viscosity once per quadrature point, and fills the two
$18\times18$ element matrices $A_e$ entry by entry,
\begin{multline*}
  (A_e)_{ij} \mathrel{+}= w_q\,|\det J|\,\eta_q
  \Bigl[\, 2\,\bm{\varepsilon}(\bm{\varphi}_i)\big|_q : \bm{\varepsilon}(\bm{\varphi}_j)\big|_q \\
  -\; \tfrac{2}{3}\,\mathrm{div}\,\bm{\varphi}_i\big|_q\;\mathrm{div}\,\bm{\varphi}_j\big|_q \,\Bigr].
\end{multline*}
In a next step, the local matrix is applied to the gathered source DoFs, $\bm{t} = A_e\,\bm{s}$ per wedge, where
$\bm{s}$ collects the eighteen wedge-local source DoFs, six nodes times three
velocity components. The result is scattered into the global destination with
$24$ atomic additions. One-thread-per-hex-cell makes for a straight-forward mapping of hexes to threads, while the split of each hex into two
wedges is what admits the reduced integration scheme of a single quadrature
point per wedge exploited in \Cref{sec:opts}.

\begin{lstlisting}[
  caption={Textbook\,\protect\markCircle{} baseline for the viscous operator, with the assembly body elided. \Cref{sec:opts} turns it into the production kernel of \Cref{lst:tuned}.},
  label={lst:kernel},
  basicstyle=\ttfamily\scriptsize,
]
struct ViscousOperator_Textbook {
 Grid4DDataVec<real,3>  src_, dst_;
 Grid4DDataScalar<real> k_;   // nodal viscosity

 void apply() {
  Kokkos::parallel_for(
   MDRangePolicy<Rank<4>>({0,..}, {n_sd,nx,ny,nr}), *this);
  Kokkos::fence();
  mpi_halo_exchange(dst_);
 }

 KOKKOS_FUNCTION
 void operator()(int sd, int x, int y, int r) const {
  Mat<real,18,18> A[2] = {};   // one per wedge
  ...                          // assemble entry by entry
  Vec<real,18> s[2] = gather_src(src_, sd, x, y, r);
  Vec<real,18> t[2] = { A[0]*s[0], A[1]*s[1] };
  scatter_atomic(dst_, sd, x, y, r, t);  // 24 atomics
 }
};
\end{lstlisting}

\paragraph{Mixed Precision.}
\label{sec:precision}
Every finite element vector in \projectname{} is precision-templated on its
scalar type, and the whole solver structure can be instantiated in reduced precision.
\projectname{} uses this to store the Krylov basis of the FGMRES restart cycle
and the multigrid preconditioner in low precision. The leading precision,
however, has to stay high: at the mesh resolutions targeted here, the round-off
error of a lower format 
would dominate the discretization error, so that the asymptotic accuracy is no longer reached
~\citep{tamstorf2021discretization,bauer2025multigrid}.
We therefore only consider double precision throughout this paper.

\section{Domain-Specific Optimisations}
\label{sec:opts}

\begin{lstlisting}[
  caption={Production\,\protect\markStar{} MV kernel, host-side launch and team body. One team owns a $t_x\!\times\!t_y\!\times\!t_r$ tile of hex cells and stages its geometry, source and coefficient DoFs into scratch behind a single barrier. Per wedge the geometry is built from the one lateral coordinate plane, since only $r_{\mathrm{mid}}$ and $\Delta r/2$ change radially, and a single centroidal quadrature point collapses the element integral.},
  label={lst:tuned},
  float=*t,
  multicols=2,
  basicstyle=\ttfamily\scriptsize,
  numbers=left,
  numberstyle=\tiny\color{gray},
  numbersep=5pt,
  xleftmargin=13pt,
  escapeinside={(*}{*)},
]
void apply() {                  // host side
 team_size = t_x * t_y * t_r;   // one thread per hex cell(*\label{ln:tile}*)
 league    = n_sd * lat_tiles*lat_tiles * r_tiles;(*\label{ln:league}*)
 TeamPolicy<LaunchBounds<maxT,minB>> policy(league,(*\label{ln:lb}*)
                    team_size);
 policy.set_scratch_size(0, PerTeam(tile_bytes));(*\label{ln:scratch}*)
 Kokkos::parallel_for(policy, *this);
 Kokkos::fence();
 mpi_halo_exchange(dst_);
}

KOKKOS_FUNCTION operator()(team) {
 // work-item identification: team -> tile,
 // thread -> one hex cell of that tile
 (sd,x0,y0,r0) = tile_origin(team.league_rank());(*\label{ln:decode}*)
 (tx,ty,tr)    = cell_in_tile(team.team_rank());(*\label{ln:decodeend}*)
 // this thread's hex cell, global indices:
 x_cell = x0 + tx;  y_cell = y0 + ty;
 r_cell = r0 + tr;  // cell = (sd,x_cell,y_cell,r_cell)

 // stage the whole tile once: HBM -> scratch.(*\label{ln:stage}*)
 par_for(TeamThreadRange(team, n_xy), [&](n) {
   // n -> (dx,dy): lateral node within the tile
   coords_sh(n,:) = grid_(sd, x0+dx, y0+dy, :); });
 par_for(TeamThreadRange(team, n_lev), [&](l) {
  // l from flat range: radial indices
   r_sh(l) = radii_(sd, r0+l); });
 par_for(TeamThreadRange(team, n_xy*n_lev), [&](t) {(*\label{ln:coalesce}*)
   src_sh(n,:,l) = src_(sd, x0+dx,y0+dy,r0+l, :);
   // could also read coefficient k from a W_0 wedge-constant function instead of W_1
   k_sh(n,l)     = k_(sd, x0+dx,y0+dy,r0+l); });
 team_barrier();          // the only barrier(*\label{ln:barrier}*)

 for (w = 0; w < 2; ++w) {         // 2 wedges
   // one centroidal q-point; only r_mid, dr vary
  // radially, the lateral plane is reused
  J = jacobian(coords_sh(tx,ty)(*\label{ln:quad}*), r_sh(tr), r_sh(tr+1));
  invJ  = inverse(J);
  // one viscosity per wedge: arithmetic, geometric
  // or harmonic mean of its 6 corner values, or a
  // constant coefficient read per wedge
  kwJ = k_eval(k_sh) * |det J|;

  // gather: strain rate eps_u(di,dj) =
  // 0.5*(d_dj u_di + d_di u_dj) and div_u at the
  // q-point, accumulated over the 18 wedge source
  // DoFs (node n, component di)
  eps_u(:,:) = 0;  div_u = 0;
  for (n = 0; n < 6; ++n) {    // wedge nodes(*\label{ln:gather}*)
   // shape gradient, reference -> physical
   grad_phi = transpose(invJ) * grad_reference[n];
   // node_of/lvl_of: wedge-local node n -> its
   // (lateral node, level) scratch slot
   for (di = 0; di < 3; ++di) {   // velocity comp
    u_di = src_sh(node_of(n), di, lvl_of(n));
    for (dj = 0; dj < 3; ++dj)  {  // derivative dir
     u_dj = src_sh(node_of(n), dj, lvl_of(n)); 
     eps_u(di,dj) += 0.5 * ( u_di * grad_phi(dj) + u_dj * grad_phi(di) );
    }
    div_u += u_di * grad_phi(di);
   }
  }


  // scatter: test eps_u and div_u against each of
  // the 18 wedge test functions (node n, comp dj)
  // and add the residuals to the global vector;
  // atomic, since neighbouring cells share nodes
  for (n = 0; n < 6; ++n) {
   grad_phi = transpose(invJ) * grad_reference[n];
   for (di = 0; di < 3; ++di) {  // 36 atomics / cell
    res = -2.0/3.0 * grad_phi(di) * div_u;
    for (dj = 0; dj < 3; ++dj)   // 2 eps(u):eps(phi)
     res += 2.0 * eps_u(di,dj) * grad_phi(dj);
    // (ddx,ddy,ddr): corner offsets of wedge node
    // n within this thread's hex 
    atomic_add(&dst_(sd, x_cell+ddx, y_cell+ddy,(*\label{ln:atomic}*)
                     r_cell+ddr, di), kwJ * res);
   }
  }
 }
}
\end{lstlisting}

In this section we lay out the domain-specific optimizations that the TERRA-NG kernels exploit.
We denote the hex-cell counts per rank by $(n_{sd}, n_x, n_y, n_r)$.
The path from the textbook baseline (Listing~\ref{lst:kernel}) to the production MV kernel is structured as the sequence of operator variants \texttt{textbook}\,\markCircle{} $\to$ \texttt{arreduc}\,\markSquare{} $\to$ \texttt{shmem}\,\markDiamond{} $\to$ \texttt{tuned}\,\markStar{} listed in \Cref{tab:opts}. \Cref{lst:tuned} is the production MV kernel that this section arrives at. 

\begin{table}[h]
\centering
\footnotesize
\setlength{\tabcolsep}{6pt}
\caption{Operator variants and the kernel optimisation each one introduces.}
\label{tab:opts}
\begin{tabular}{clp{0.62\linewidth}}
\toprule
phase & variant & optimisation \\
\midrule
\markCircle  & \texttt{textbook} & naive, textbook-style matrix-free MV \\
\markSquare  & \texttt{arreduc}   & single quadrature point per wedge, code-generated element kernels, dimension-wise fused assemble-and-apply \\
\markDiamond & \texttt{shmem}    & hierarchical parallelism, shared memory, and global memory coalescing \\
\markStar    & \texttt{tuned}    & register and occupancy tuning to the arch at hand \\
\bottomrule
\end{tabular}
\end{table}

\paragraph{\texttt{arreduc}\,\protect\markSquare: Single Quadrature Point per Wedge.}
Mantle viscosity $\eta$ varies by four to six orders of magnitude across the thin boundary layer at the lithosphere surface boundary. The established geodynamics response is one effective viscosity per hex cell, either as a centroidal evaluation or as an arithmetic, harmonic, or geometric per-hex-cell mean, because multi-point sampling across such a near-discontinuity produces non-physical pressure artefacts~\citep{Deubelbeiss:2008:PEPI,Heister:2017:GJI}. The finite-volume codes StagYY~\citep{tackley2008} and CitcomS~\citep{zhong2000,zhong2008} build this directly into the discretisation. The wedge split inherited from TERRA's original icosahedral discretisation~\citep{baumgardner1985icos} enables carrying a reduced-integration regime into a finite-element setting with a similar effect: a single centroidal quadrature point per wedge, requiring far less arithmetic than the standard six-point rules and evaluating the viscosity once. Its accuracy and stability are analysed in the companion paper~\citep{boehm2026terrang}, while we focus on its impact on performance compared to a full, 6 point quadrature scheme here. This is a change of discretisation, not only of implementation: the rule truncates the mixed bilinear terms and admits element-local hourglass modes. It is not a change of accuracy: the optimal discretisation order is retained, and hourglass control restores textbook multigrid convergence~\citep{boehm2026terrang}. The collapsed inner quadrature loop reduces to one Jacobian, one $\eta$ lookup, and one set of basis gradients per wedge, with the small negligible overhead of the hourglass control. In \Cref{lst:tuned} this is line~\ref{ln:quad}: the geometry is evaluated once per wedge, at the centroid and the mid-radius, and no quadrature loop remains.

\begin{figure*}[t]
  \centering
  \raisebox{-0.5\height}{\includegraphics[width=0.47\textwidth]{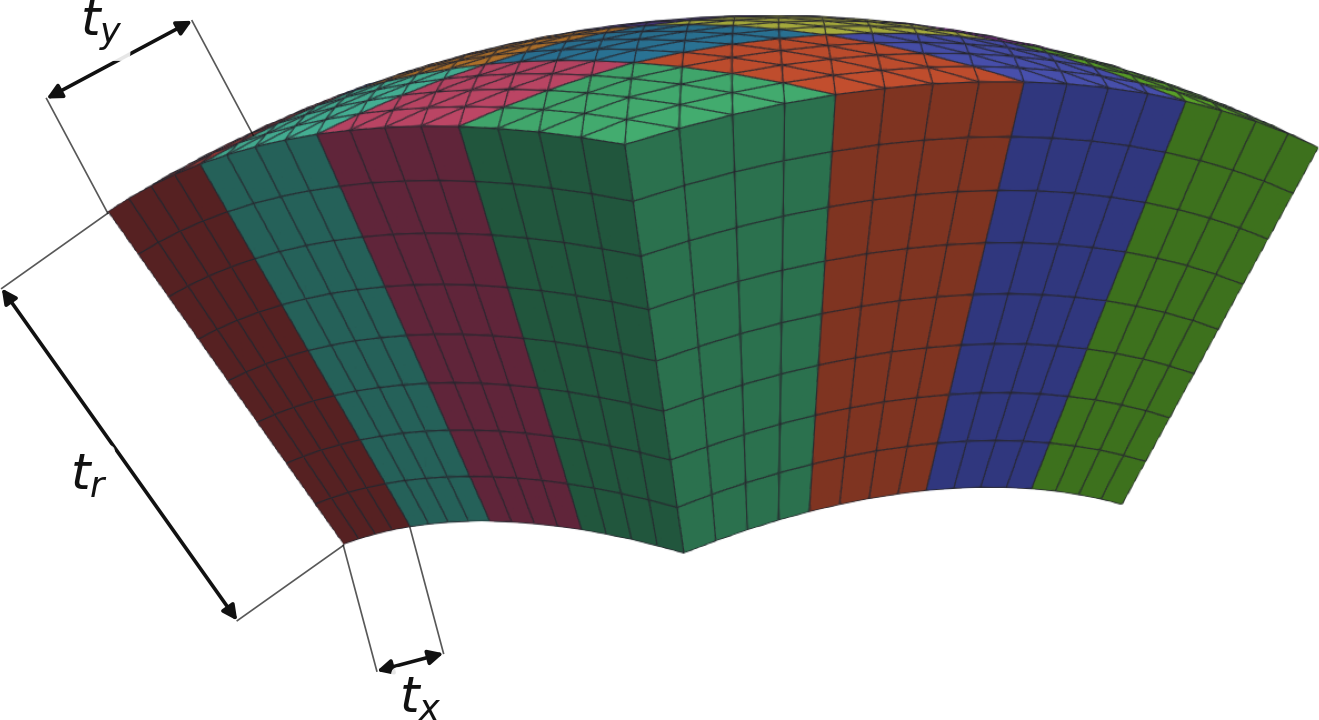}}%
  \hfill
  \raisebox{-0.5\height}{\includegraphics[width=0.47\textwidth]{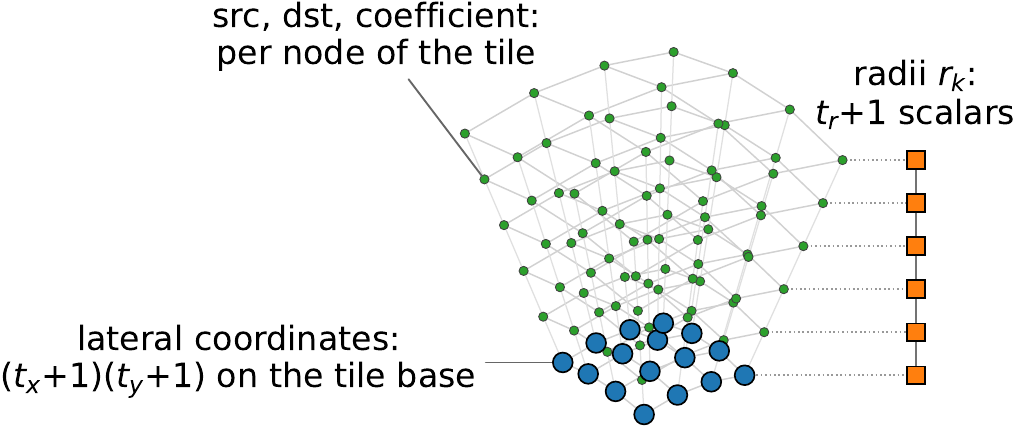}}
  \caption{Thread-team tiling and the per-tile scratch data. Left:
  one subdomain viewed from the side. Each colour marks one team, which
  owns a $t_x \times t_y \times t_r$ tile of hex cells (here
  $4\times4\times8$, spanning the full radial extent) and processes them
  with one thread per hex cell. Right: the data each team stages into
  on-chip scratch memory for its tile: the $(t_x{+}1)(t_y{+}1)$
  lateral coordinates of the tile base (blue), from which each thread
  extrudes the geometry on-the-fly using the $t_r{+}1$ per-layer radii
  (orange), and the source and coefficient values on every
  node of the tile (green).}
  \label{fig:teams}
\end{figure*}

\begin{figure}[t]
  \centering
  \includegraphics[width=\columnwidth]{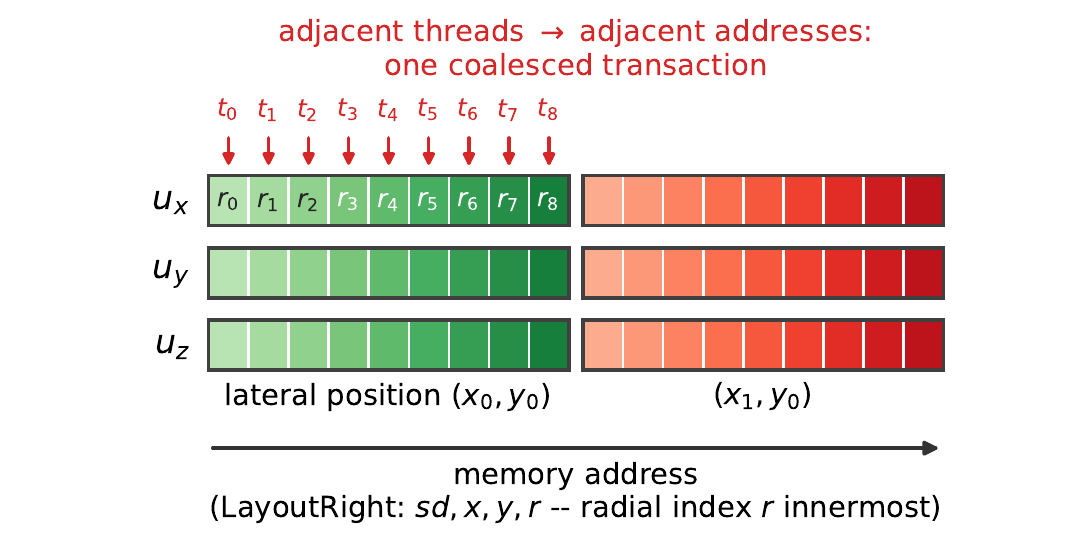}
  \caption{Memory layout of the grid data for a shell of only 8 radial cells, so 9 radial DoFs per lateral position. Each velocity component is one
  contiguous \texttt{Grid4DDataScalar} View. Within
  each View, \texttt{Kokkos::LayoutRight} places the radial index $r$
  innermost, so consecutive radial entries of one lateral position are
  contiguous in memory. Adjacent thread IDs of the cooperative
  \texttt{TeamThreadRange} load therefore hit adjacent addresses.
}
  \label{fig:memlayout}
\end{figure}

\paragraph{\texttt{arreduc}\,\protect\markSquare: SymPy-based Kernel Generation.}
The element-level arithmetic of every operator is produced by a SymPy-based code generator, following the approach of the HyTeG Operator Generator~\citep{bohm2025code}, that symbolically integrates the weak form for the chosen quadrature rule, applies source-level common-subexpression elimination, and emits scalar C++ expressions for the fused gather--scatter kernel. Geometry-dependent terms (Jacobians, determinants) are factored out where possible, removing hand-written element-level code, minimizing redundant floating-point operations, and making it cheap to add new operators.  This is what underlies the gather and scatter loops of \Cref{lst:tuned}, lines~\ref{ln:gather}--\ref{ln:atomic}, although the low-level arithmetic is omitted in the pseudocode.

\paragraph{\texttt{arreduc}\,\protect\markSquare: Dimension-wise Fused Assemble-and-Apply.}
The production kernel never materializes the local element matrix $A_e$ like the textbook, which would require $n_\text{quad} \cdot d^2 \cdot n_\text{w-dofs}^2 = n_\text{quad} \cdot 324$ operations per wedge.  Instead, it fuses the assembly of the local matrix and its matrix-vector multiplication. The resulting gather and scatter phases reduce the complexity to $n_\text{quad} \cdot d^2 \cdot 2 \cdot n_\text{w-dofs} = n_\text{quad} \cdot 108$. This fused assembly-and-application is a standard technique in matrix-free FE computations~\citep{Kronbichler:2012:CAF}. In contrast to precomputing and storing the per-quadrature-point Jacobians, our kernels recompute the geometry on-the-fly from the radially compressed coordinate storage, so no per-element geometry data is streamed from main memory. 

 The same folding can be applied on the block-level of the $d \times d$-vectorial viscous operator: treating the $3 \times n_\text{w-dofs}$ shape functions as one set of vectorial shape functions coupled by a monolithic viscous operator instead of three sets of scalar shape functions coupled by a $d \times d$-block operator gives the theoretical justification to also fold the nested dimensions loop, reducing complexity further to $n_\text{quad} \cdot 2 \cdot d \cdot 2 \cdot n_\text{w-dofs} = n_\text{quad} \cdot 72$. The two loops of \Cref{lst:tuned} at lines~\ref{ln:gather} and~\ref{ln:atomic} are this fusion: the first accumulates the source into the strain rate for all 6 source DoFs per wedge and 3 velocity components. The second loop contracts the accumulated strain rate against every test gradient into the global destination.

\paragraph{\texttt{shmem}\,\protect\markDiamond: Hierarchical Parallelism and Shared Memory.}
Due to the continuous Galerkin discretization, every nodal source, coefficient and destination DoF is shared by up to eight hex cells (four lateral neighbours times two radial layers) and gathered once per hex-cell thread. A naive cell-per-thread or wedge-per-thread dispatch therefore loads the same data mostly redundantly from main memory, once per visiting thread.

We exploit the second level of Kokkos' parallelism, \texttt{TeamPolicy}, to facilitate staging shared DoFs in shared memory. A team owns a tile of $t_x \times t_y \times t_r$ hex cells (\Cref{fig:teams}) and contains exactly $N_\mathcal{T} = t_x\,t_y\,t_r$ threads, one per hex cell. With the per-rank hex-cell counts $(n_{sd}, n_x, n_y, n_r)$ from Section~\ref{sec:grid}, the number of teams launched per rank is
\begin{equation*}
  n_\mathcal{T} \;=\; n_{sd}\;\bigl\lceil n_x / t_x \bigr\rceil\;\bigl\lceil n_y / t_y \bigr\rceil\;\bigl\lceil n_r / t_r \bigr\rceil,
\end{equation*}
so the grid- and team-level dispatches together cover every hex cell exactly once. In case the global dimensions of the compute grid are not divisible cleanly by the team size in each dimension, threads are masked. In \Cref{lst:tuned} the team-based dispatch is set up at lines~\ref{ln:tile}--\ref{ln:league} and each thread decodes its tile origin and its hex cell within the tile at lines~\ref{ln:decode} and~\ref{ln:decodeend}. Before any thread enters the per-hex-cell kernel, the team cooperatively stages the tile's shared data into Kokkos scratch memory using \texttt{TeamThreadRange} parallel-fors (lines~\ref{ln:stage}--\ref{ln:barrier}), closed by the single barrier. This includes $n_{xy} = (t_x+1)(t_y+1)$ lateral coordinates (which are later radially extruded on-the-fly by each thread), $n_{lev} = t_r + 1$ for the radii, and most importantly, the $4 \cdot n_{xy} \cdot n_{lev}$ nodal source and coefficient DoFs. A single \texttt{team\_barrier()} follows, after which every geometry, source or coefficient access inside the kernel hits on-chip scratch rather than HBM. The tile dimensions are the 
central means to tune the code: $N_\mathcal{T}$ must fit the vendor's max threads-per-block, and the scratch footprint must fit the available on-chip memory per SM/CU/Xe-core. Neglecting DoFs on tile-boundaries, this reduces the amount of global loads from $8$ to $1$ per source and coefficient DoF.

\paragraph{\texttt{shmem}\,\protect\markDiamond: Global Memory Coalescing.}
After the cooperative staging just described, no per-hex-cell access inside the kernel reaches HBM. The coalescing question therefore moves to the \texttt{TeamThreadRange} loads that fill the scratch in the first place (\Cref{fig:memlayout}). Kokkos defaults to \texttt{LayoutLeft} on its GPU backends, which would place the subdomain index, a virtually unused axis with $n_{sd} \sim \mathcal{O}(1)$ entries per rank, contiguous in memory and stride the radial accesses our kernel actually walks. We instead instantiate every grid data View explicitly with \texttt{Kokkos::LayoutRight} so that the radial index $r$ is innermost and adjacent radial entries are contiguous. Vector quantities are stored as structure-of-arrays via \texttt{Grid4DDataVec}: each velocity component occupies its own \texttt{Grid4DDataScalar} View, and a thread loading all three components issues three independently contiguous transactions rather than one dimension-strided one. Inside the cooperative load, $\mathtt{TeamThreadRange}(\mathrm{team}, n_{xy}\!\cdot\!n_{lev})$ linearises the team's threads over the (lateral-node, radial-level) pairs of the staged tile, with the radial level changing fastest, adjacent thread IDs hit adjacent addresses and every HBM transaction is fully coalesced across the warp, wavefront, or subgroup.

\paragraph{\texttt{tuned}\,\protect\markStar: Architecture-Specific Tuning.}
Starting from the architecture-agnostic \texttt{shmem}\,\markDiamond{} kernel, the \texttt{tuned}\,\markStar{} variant applies per-vendor adjustments. Note that this does not include vendor-specific instructions: all variants can run on all GPUs, as they are purely using C++ and Kokkos language constructs. In that sense also \texttt{tuned}\,\markStar{} is architecture-agnostic. However, there are certain characteristics of the kernel, e.g. how many registers, how much scratch memory it requires and the shape of the tiles, that are better on one architecture or another, and
that can be optimized for the GPU at hand. 
\paragraph{\texttt{tuned}\,\protect\markStar: Tile-size selection.} The team-tile dimensions $(t_x, t_y, t_r)$ set the team size and the scratch footprint. They are swept per architecture and pinned to the value that maximises measured throughput, subject to the threads-per-block and on-chip-scratch limits of each vendor (NVIDIA SMs, AMD CUs, Intel Xe-cores). \Cref{sec:tileshape} reports that sweep. They enter \Cref{lst:tuned} at lines~\ref{ln:tile} and~\ref{ln:scratch}.
\paragraph{\texttt{tuned}\,\protect\markStar: Cache vs.\ recompute trade-off, Occupancy and Atomics.} Where \texttt{shmem}\,\markDiamond{} caches intermediate quantities (Jacobians, basis gradients, geometry-derived terms) in registers or scratch, \texttt{tuned}\,\markStar{} selectively recomputes them on the fly to free registers and potentially hit a higher occupancy bin. In some cases, accumulators for e.g. the destination DoFs, which increases local memory, decreases occupancy but e.g. reduces number of atomics and arithmetic can also be advantageous. Furthermore, launch bounds let the programmer influence registers per thread against achieved occupancy manually. This is layed out in \cref{sec:acc} and \cref{sec:launchbounds}.

\section{Cross-architecture Comparison}
\label{sec:cross-vendor}
In this section we compare how the optimised kernel performs across the four GPU architectures, where it sits relative to each device's roofline, and what the remaining tuning space looks like on each of them.

\begin{table}[h]
\centering
\footnotesize
\setlength{\tabcolsep}{5pt}
\caption{Compiler and MPI used on each machine. The same source is compiled with each vendor's own toolchain.}
\label{tab:environments}
\begin{tabular*}{\columnwidth}{@{\extracolsep{\fill}}lll@{}}
\toprule
machine & compiler & MPI \\
\midrule
MareNostrum~5 & NVHPC 23.11     & Open MPI (HPC-X 2.16) \\
LUMI-G        & ROCm 6.3.4      & Cray MPICH 8.1.32 \\
Hunter        & ROCm 6.4.1      & Cray MPICH 8.1.33 \\
SuperMUC-NG~2 & oneAPI 2025.2.0 & Intel MPI 2021.16 \\
\bottomrule
\end{tabular*}
\end{table}

We profile with \texttt{Nsight} \texttt{Compute} on the H100, \texttt{rocprofv3} on MI250X and MI300A, and \texttt{Intel} \texttt{Oneprof} on PVC. \Cref{tab:environments} lists the compiler used on each machine.

\Cref{tab:gpus} summarises the per-device peak specifications of the four GPUs, as deployed in the respective host supercomputer, that the rooflines, throughput, and bandwidth figures of this section are placed against. For MI250X and PVC we report the per-GCD / per-tile values since these are the units the Kokkos backend addresses (the OAM module / package carries two of each).

\begin{table}[h]
\centering
\footnotesize
\setlength{\tabcolsep}{5pt}
\caption{Per-device peak specifications of the four GPUs in their host supercomputer configuration. ``scratch'' is the on-chip programmer-managed memory per SM / CU / Xe-core (shared memory, LDS, SLM). ``L2'' is the device-level L2 cache (for MI300A the aggregate across the six XCDs, not counting the additional 256\,MB Infinity Cache).}
\label{tab:gpus}
\resizebox{\columnwidth}{!}{%
\begin{tabular}{lccccc}
\toprule
device & HBM & BW [TB/s] & scratch [KB] & L2 [MB] & FP64 [TF] \\
\midrule
H100 SXM5       & 64\,GB         & 2.68  & 228 & 50   & 33.5 \\
MI250X / GCD     & 64\,GB       & 1.6   & 64  & 8    & 23.9 \\
MI300A APU      & 128\,GB & 5.3   & 64  & 24   & 61.3 \\
PVC 1550 / tile & 64\,GB         & 1.638 & 128 & 204  & 26.0 \\
\bottomrule
\end{tabular}}
\end{table}

\subsection{Roofline Placement}
\label{sec:roofline}

\begin{figure}[t]
  \centering
  \includegraphics[width=\columnwidth]{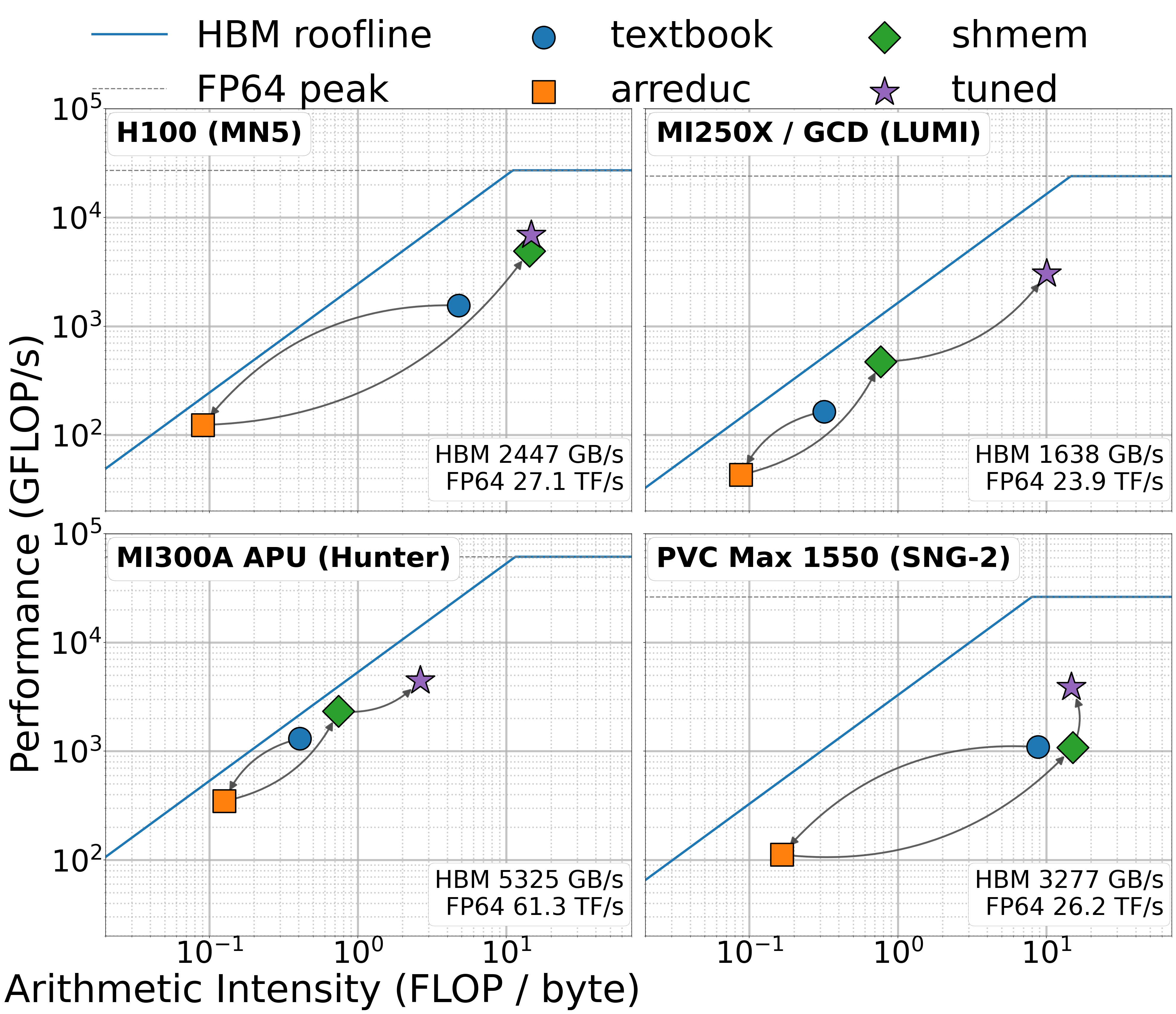}
  \caption{Main-memory (HBM) roofline of \projectname{}'s matrix-free MV kernel on H100 (MareNostrum~5), MI250X per GCD (LUMI-G), MI300A APU (Hunter), and PVC Max~1550 (SuperMUC-NG Phase~2). The four markers correspond to the variants of \Cref{tab:opts}. Arrows indicate the optimisation trajectory \texttt{textbook}\,\protect\markCircle{} $\to$ \texttt{arreduc}\,\protect\markSquare{} $\to$ \texttt{shmem}\,\protect\markDiamond{} $\to$ \texttt{tuned}\,\protect\markStar{}.}
  \label{fig:rooflines-hbm}
\end{figure}

We first characterize the kernel and how our optimizations impact it with respect to speed-of-light measures in the roofline model.
\Cref{fig:rooflines-hbm} places each of the four operator variants on the main-memory (HBM) roofline of H100, MI250X (per GCD), MI300A, and PVC. The four markers from \Cref{tab:opts} thread through the same trajectory on every architecture: \texttt{textbook}\,\markCircle{} sits far below the roof on most GPUs, \texttt{arreduc}\,\markSquare{} reduces (redundant) arithmetic, dropping the numerator of the arithmetic intensity, thereby pulling the operating point left while raising throughput, \texttt{shmem}\,\markDiamond{} eliminates redundant main-memory accesses by staging the shared DoFs in scratch memory, increasing the arithmetic intensity by reducing its denominator and climbing along the bandwidth ceiling, and \texttt{tuned}\,\markStar{} gives depending on the GPU another upwards boost and 
achieves a performance near the HBM knee.

We observe that the same qualitative pattern holds on all four architectures, testifying some universality of the optimizations. However, the contribution of each phase varies vastly between vendors: the final tuning step, for example, \texttt{tuned} \markStar{} improves performance significantly on MI250X and PVC but only to a smaller extent on the H100. This might stem from a more mature compiler environment on NVIDIA, or from the AMD and Intel backends of Kokkos being less thoroughly optimised. In either case the NVIDIA toolchain already makes near-optimal choices for the tile size and for caching versus recomputing intermediate variables (the techniques described in \cref{sec:opts}, \texttt{tuned} \markStar{}), leaving manual trial-and-error tuning less to recover on NVIDIA and correspondingly more on the other architectures.
 
\subsection{Throughput}
\label{sec:throughput}

\begin{figure}[t]
  \centering
  \includegraphics[width=\columnwidth]{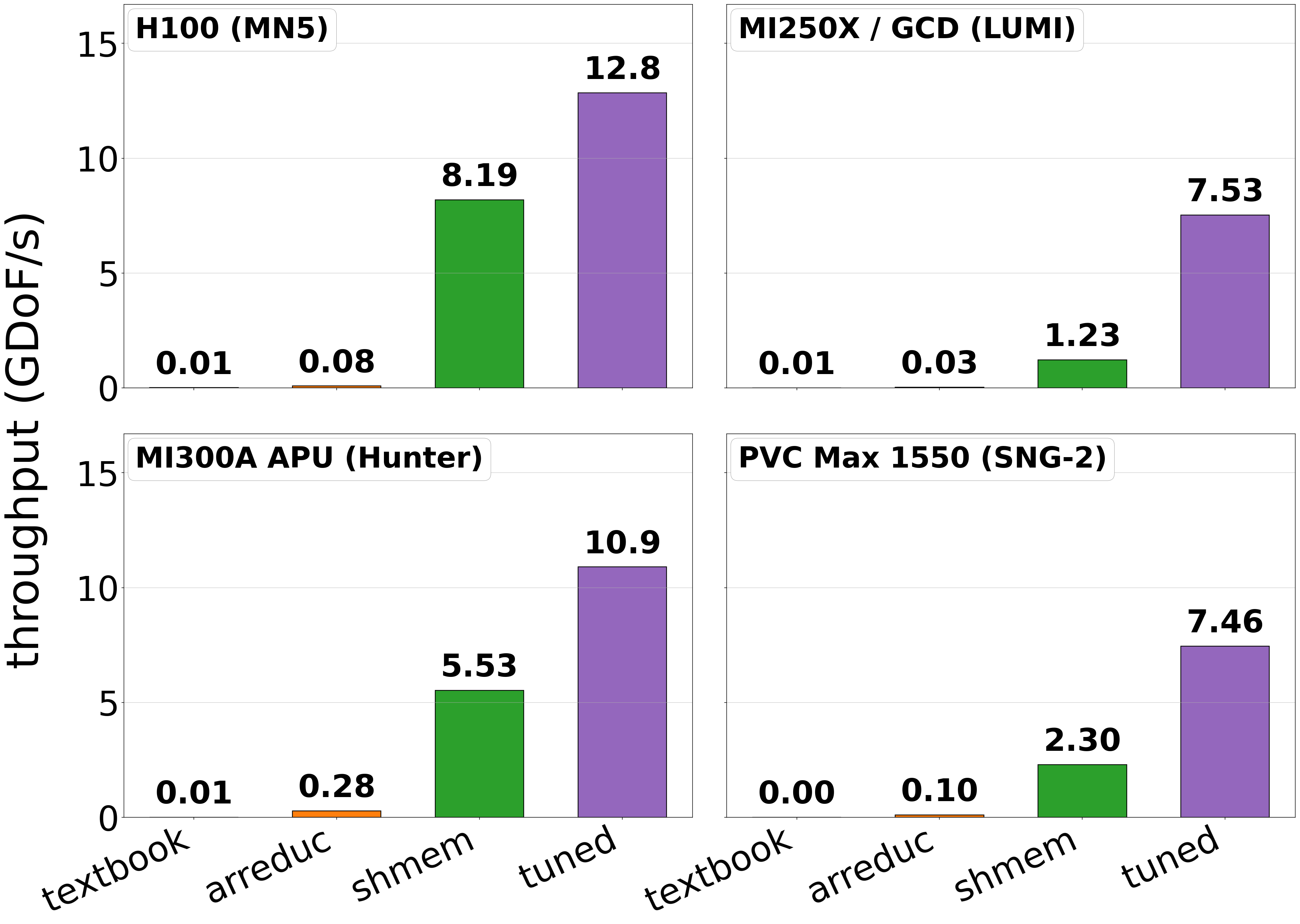}
  \caption{Throughput (GDoF/s) on a single GPU of the matrix-free MV kernel of \Cref{lst:kernel} together with the optimizations from \Cref{sec:opts} applied. MI300A remains limited by the atomic scatter analysed in \Cref{sec:acc}.}
  \label{fig:phase-throughput}
\end{figure}

The improved FLOPs performance and the shift towards the HBM roofline shown in the roofline plots \Cref{fig:rooflines-hbm} are not pure performance improvements, but translate directly into gains on the ultimately relevant metric, operator throughput. Throughput here is the number of velocity DoFs the operator updates per second, $N_{\mathrm{vel}}/t_{\mathrm{apply}}$, with $N_{\mathrm{vel}}$ the three velocity components on every node of the level-$\ell$ mesh and $t_{\mathrm{apply}}$ the wall time of a single matrix-free apply on one device. GDoF/s denotes $10^{9}$ of them per second.

\Cref{fig:phase-throughput} reports the throughput of the matrix-free apply  in GDoF/s across the same four-phase progression: on every architecture the throughput rises monotonically through \texttt{textbook}\,\markCircle{} $\to$ \texttt{arreduc}\,\markSquare{} $\to$ \texttt{shmem}\,\markDiamond{} $\to$ \texttt{tuned}\,\markStar{}. The relative gains mirror the roofline picture: the final tuning step buys only $1.6\times$ on H100, against $2.0\times$ on MI300A, $3.2\times$ on PVC and $6.1\times$ on MI250X. Our code nonetheless runs fastest on the H100, and sometimes already before that step: at \texttt{shmem}\,\markDiamond{} the H100 delivers more throughput than MI250X and PVC reach even after tuning. Conversely, and mirroring its distance from the roof in \Cref{fig:rooflines-hbm}, the MI300A stays comparatively slow at $10.9$\,GDoF/s although it is the newest and, on paper, the most capable device considered, with roughly twice the HBM bandwidth and peak FP64 rate of the H100 (\Cref{tab:gpus}). The kernel is appearently limited by other parts of the GPU than those top level specs.

\begin{figure}[t]
  \centering
  \includegraphics[width=\columnwidth]{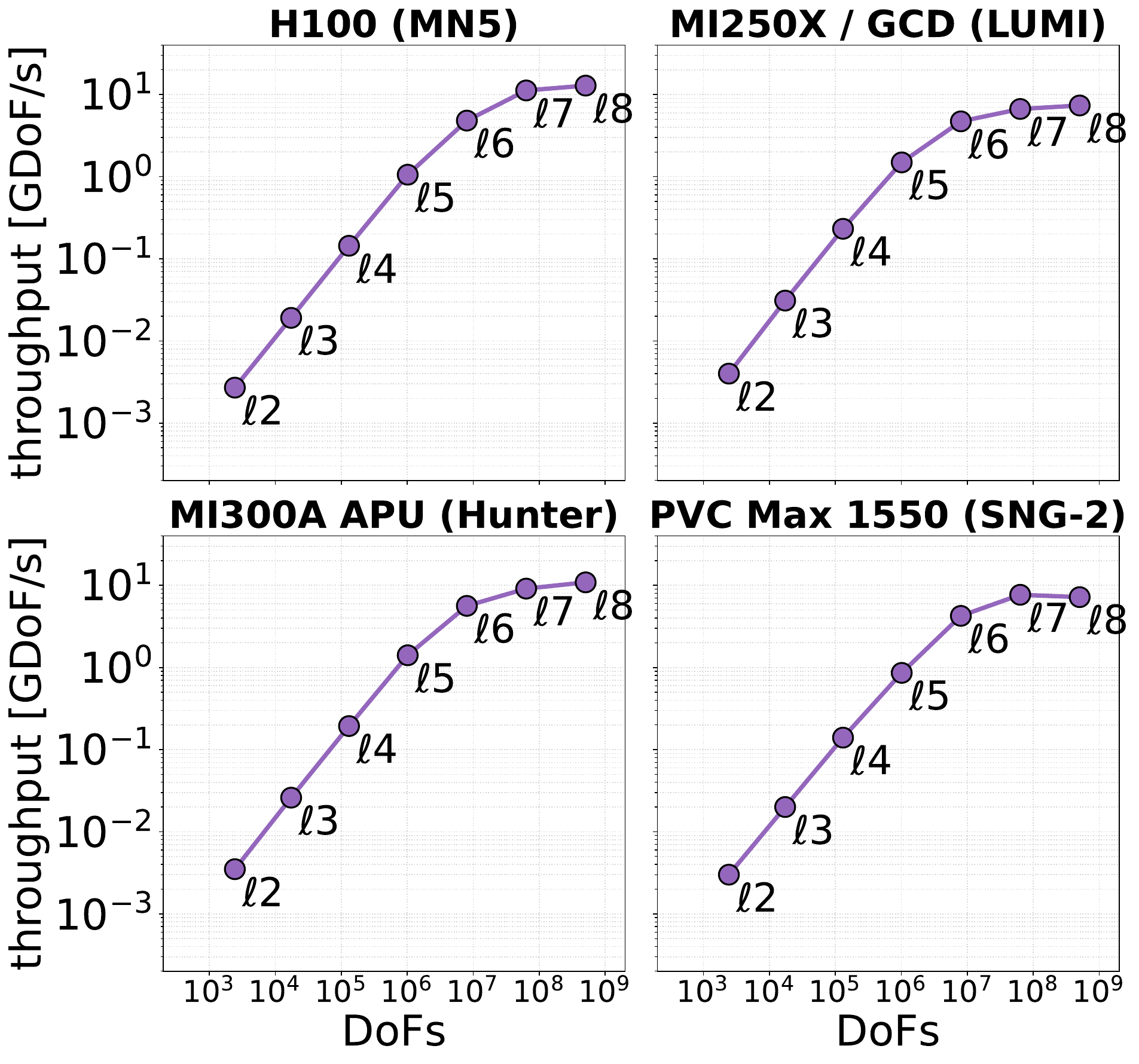}
  \caption{Throughput of the \texttt{tuned}\,\protect\markStar{} MV kernel as a function of problem size, sweeping the refinement level $\ell$ on a single device.
  }
  \label{fig:throughput-level}
\end{figure}

\Cref{fig:throughput-level} sweeps the refinement level on a single device and shows that the peak throughput is reached only once the device is saturated. Up to $\ell5$ the grid is too small to fill the machine and the measurement is dominated by kernel-launch and latency overheads rather than by memory traffic. Saturation sets in between $\ell6$ and $\ell7$, and beyond $\ell7$ the curves flatten into a plateau. The plateau value is the per-device asymptote used in \cref{fig:phase-throughput}, and only levels in this regime are representative of production runs, where every device holds a saturating share of the global grid.

\subsection{\texttt{Tuned}\protect\markStar\ : Team-Tile Shape}
\label{sec:tileshape}
\begin{figure}[t]
  \centering
  \includegraphics[width=\columnwidth]{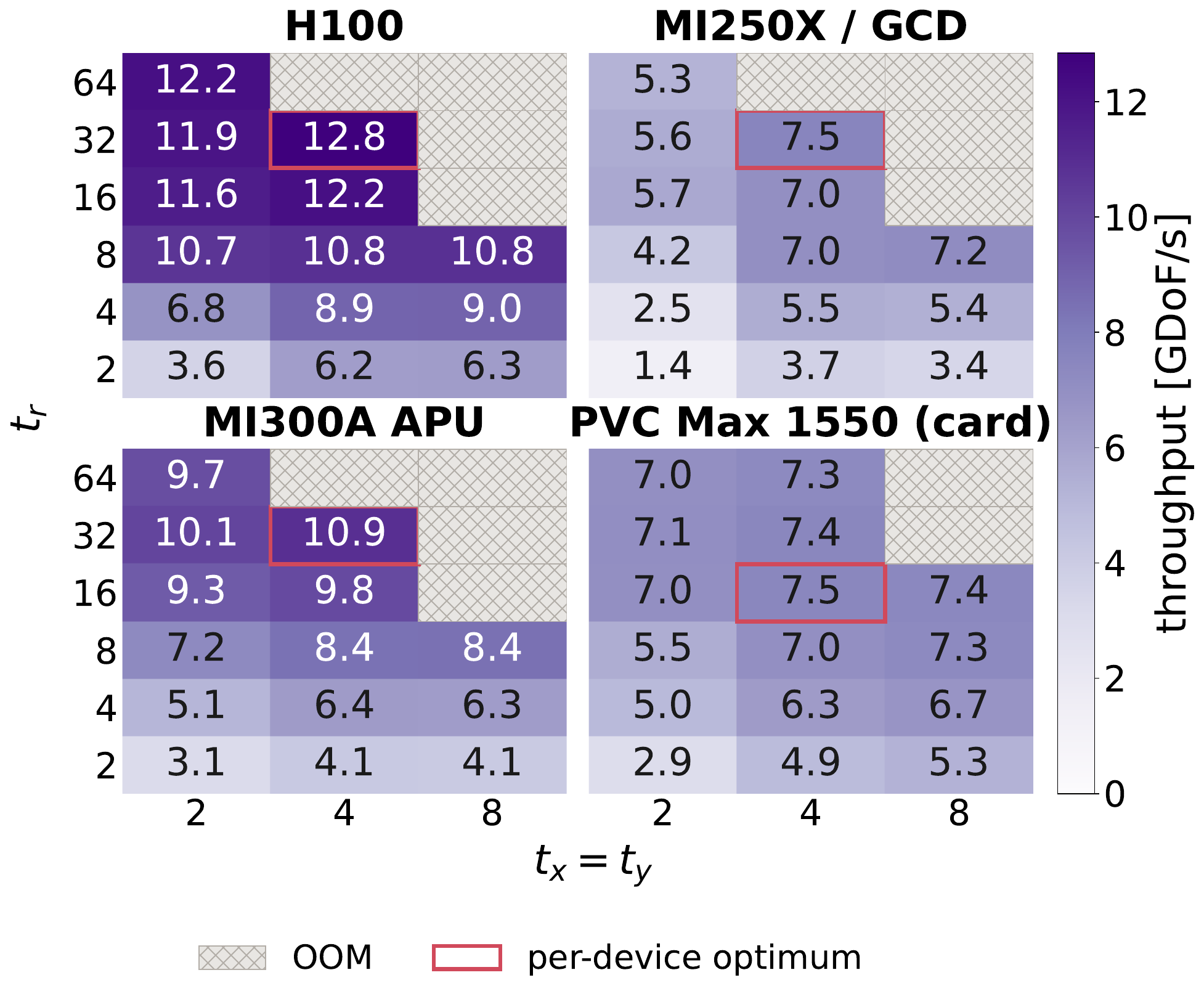}
  \caption{Throughput of the \texttt{tuned}\,\protect\markStar{} MV kernel over the team-tile parameter space, on a single device at $\ell8$ ($2.54\times10^{8}$ velocity DoFs). Hatched cells could not be run: they exceed the device's work-group size or scratch-memory limit.}
  \label{fig:tile-sweep}
\end{figure}

We investigate how far we can push the throughput by finding optimal team-tile sizes $t_x, t_y, t_r$. Kokkos exposes the overall team-size but how the team is spread laterally and radially is on the user. \Cref{fig:tile-sweep} sweeps this space exhaustively at $\ell8$.

The optimum is the same on three of the four devices: H100 and both AMD parts peak at $t_x=t_y=4$, $t_r=32$, and only PVC prefers a 
tile with dimensions closer to a square, $t_r=16$. Throughput varies by $5.5\times$ across the feasible tiles on MI250X, $3.6\times$ on H100, $3.5\times$ on MI300A and $2.5\times$ on PVC. 
The two 
extremes of the range fail for different reasons. 
Large teams exhaust the scratch memory and do not launch at all (hatched cells in \Cref{fig:tile-sweep}), while small teams run but amortise the cost of team administration, the cooperative loads, the barriers, and the halo staged once per team, over too little work, such that the smallest $t_r=2,t_x=t_y=2$-tile is the slowest configuration on every architecture. Aspect ratio matters as much as size: at a fixed team size a flat lateral plate is always the worst choice. This is intuitive given the $r$-fastest-moving-dimension memory layout (\cref{fig:memlayout}). Tiling mostly radially ($t_x=2$ configs) performs well close to the optimum, but a certain degree of lateral reuse seems especially beneficial on MI250X.

\subsection{Memory Utilisation}
\label{sec:memutil}
\begin{figure}[t]
  \centering
  \includegraphics[width=\columnwidth]{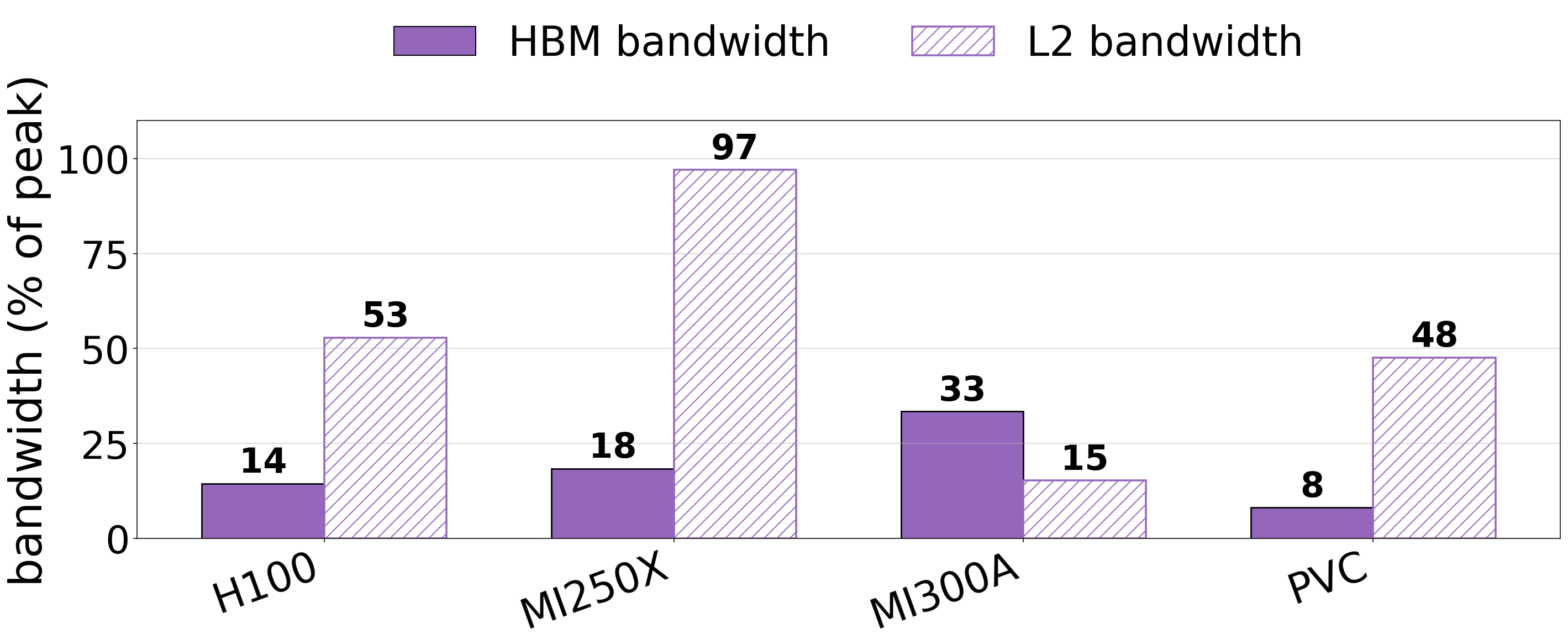}
  \caption{Achieved HBM (solid) and L2 (hatched) bandwidth of the \texttt{tuned}\,\protect\markStar{} production MV kernel as a fraction of per-vendor peak.}
  \label{fig:phase-bw}
\end{figure}

From \texttt{arreduc}\,\markSquare{} onwards the kernel has low arithmetic intensity, lower than or at the 
roofline knee-point on all GPUs, so its speed is
determined by how well it maps onto the memory hierarchy, 
including lower-level caches, rather than by how much arithmetic it performs.
\Cref{fig:phase-bw} reports the HBM and L2 bandwidth achieved by the \texttt{tuned}\,\markStar{} production MV kernel as a fraction of per-vendor peak.
The kernel sustains a high fraction of the L2 bandwidth on most architectures, while the achieved HBM bandwidth stays comparatively low. On MI300A, that balance is inverted with a higher relative memory than L2 cache bandwidth.

Using hardware counters, we observe that atomics are $87\,\%$ of all L2 requests on MI250X and $90\,\%$ on MI300A, so the L2 traffic is dominated by atomics. This is expected: the gather loading source and coefficient DoFs from main memory and scatter writing destination DoFs to main memory atomically amortise very differently over a team. The gather is staged in scratch once per team and then read only from scratch by every hex cell. The scatter cannot be amortised at all, since neighbouring teams and cells contribute to the same nodes and each cell must therefore accumulate into the global destination. Writing $n_{xy}=(t_x+1)(t_y+1)$ for the nodes of a lateral layer and $n_{r}=t_r+1$ for the radial layers, a team stages
\begin{equation}
  N_{\mathrm{gather}}
  = \underbrace{3\,n_{xy}}_{\text{coordinates}}
  + \underbrace{3\,n_{xy}n_{r}}_{\text{source}}
  + \underbrace{n_{xy}n_{r}}_{\text{viscosity}}
  + \underbrace{n_{r}}_{\text{radii}}
  \label{eq:gather}
\end{equation}
DoFs/doubles to shared memory, while it scatters  
\begin{equation}
  N_{\mathrm{scatter}}
  = \underbrace{t_x \cdot t_y \cdot t_r}_{\text{hex cells per team}}
  \cdot \underbrace{2\cdot18}_{\substack{\text{wedges}\,\times\,\text{DoFs per wedge}}}
  \label{eq:scatter}
\end{equation}
times with atomic additions. For a tile with e.g. $t_x=t_y=4$, $t_r=32$, we get $N_{\mathrm{gather}}=3408$ against $N_{\mathrm{scatter}}=18\,432$, so the operator issues roughly five atomic requests for every value it reads. 

MI300A is an outlier with comparatively high memory traffic. It is an APU whose GPU and CPU share one address space, so an atomic has to be coherent with host accesses to the same address and not merely with the six XCDs and their private L2 caches. It therefore cannot be completed in a XCD-local L2 and is resolved in the data fabric, backed by Infinity Cache. This is documented behaviour rather than an artefact of our kernel: the CDNA2 ISA states that memory atomics are performed in the last level texture cache, i.e. the MI250X GCD's L2, whereas the CDNA3 ISA states that they are performed in the data fabric and are therefore known to be atomic with host memory access (Advanced Micro Devices 2022, 2024). On MI300A, the operator therefore pays the L2 request traffic every device pays and, on top, the traffic a single-die device avoids, which causes the comparatively high memory-side bandwidth of 33 \% in \Cref{fig:phase-bw}. This is the reason for our code running slower on the MI300A than the older H100. 

\subsection{\texttt{Tuned}\protect\markStar\ : Atomics vs Local Accumulator on MI300A}
\label{sec:acc}

A remedy reduces the number of atomics: instead of atomically adding  the destination DoFs  during the scatter phase, we accumulate them in a private copy of the destination DoFs and flush with one atomic per DoF after the scatter loop. For that, two variants come to mind: first, storing the destination DoFs locally in a per-thread stack array where a thread holds only its own hex cell, whose two wedges scatter $2\cdot6\cdot3$ DoFs but share four of their six nodes and so touch only $8\cdot3$ distinct ones per hex cell, giving
\begin{equation}
  \frac{2\cdot6\cdot3}{8\cdot3} = \frac{36}{24} = \frac{3}{2}
  \label{eq:atomicreductioncell}
\end{equation}
fewer atomics, independently of the tile-size. Second, the destination DoFs for the whole tile can be held in team scratch at once, such that $N_{\mathrm{scatter}}$ collapses to the $3\,n_{xy}n_{r}$ DoFs the team actually touches, and the reduction is
\begin{equation}
  \frac{N_{\mathrm{scatter}}}{3\,n_{xy}n_{r}}
  = \frac{12\,t_x t_y t_r}{(t_x+1)(t_y+1)(t_r+1)},
  \label{eq:atomicreduction}
\end{equation}
or $18\,432\to2\,475$ atomics per team, a factor of $7.4$, for the example $t_x=t_y=4$, $t_r=32$ tile. 

In which memory that accumulator lives decides the impact on throughput, summarized in \cref{tab:scattercache}. Held per hex cell as a C stack array ("per hex cell" row in \cref{tab:scattercache}) it causes register spills of $208$\,B/lane, which drives down occupancy from 4 to 3 waves/SIMD. The per-hex-cell variant drastically tanks throughput because its spill goes to memory as well, raising the memory bandwidth drastically on both considered cards.

Staged per team tile it lives in scratch, leaves the registers untouched and pushes occupancy to only 2 waves/SIMD ("per tile" row in \cref{tab:scattercache}). That trade only pays where the atomics reach memory: MI250X resolves atomic writes in L2 cache, so reducing them does not yield throughput gain, but the reduced occupancy is a slight throughput loss. On the MI300A they go through the L2 cache, to infinity cache with much higher latency, so the reduction in atomics yields $27\,\%$ improved throughput, beating the H100 and achieving the highest throughput of all devices. Here, the same change of source results in a substantial performance gain on one architecture and a loss on another.
\begin{table}[h]
\centering
\scriptsize
\setlength{\tabcolsep}{1.5pt}
\caption{Accumulating the destination DoFs before the atomic scatter, tile size $4/4/32$. Spill memory [B/lane], scratch [KB] and occupancy [waves/SIMD] are observed. L2 and memory are bandwidths in GB/s, throughput in GDoF/s. With the team-scratch accumulator for dst DoFs, MI300A exceeds the throughput it reaches in \Cref{fig:phase-throughput}, while the same change tanks throughput on MI250X.
}
\label{tab:scattercache}
\begin{tabular*}{\columnwidth}{@{\extracolsep{\fill}}lrrrrr>{\columncolor{tblviolet}\color{white}}rrr>{\columncolor{tblviolet}\color{white}}r@{}}
\toprule
 & \multicolumn{3}{c}{both devices} & \multicolumn{3}{c}{MI250X} & \multicolumn{3}{c}{MI300A} \\
\cmidrule(lr){2-4}\cmidrule(lr){5-7}\cmidrule(lr){8-10}
accumulator & spills & scratch & occup. & L2 & mem & thrpt. & L2 & mem & thrpt. \\
\midrule
none     & $0$   & $31$ & $4$ & $972$ & $208$ & $7.54$ & $1220$ & $1180$ & $10.90$ \\
per hex cell & $208$ & $31$ & $3$ & $877$ & $651$ & $2.46$ & $2084$ & $2277$ & $6.27$ \\
per tile/team & $0$   & $50$ & $2$ & $212$ & $178$ & $6.50$ & $384$  & $503$  & $13.87$ \\
\bottomrule
\end{tabular*}
\end{table}

\subsection{\texttt{Tuned}\protect\markStar\ : Launch Bounds}
\label{sec:launchbounds}
The compiler fixes registers, spill and occupancy when it compiles the team, and the accumulator in \cref{sec:acc} only changed them indirectly, by asking for more local memory in terms of additional registers or scratch. They can also be steered directly from the source, through the launch bounds. Kokkos exposes that hint as a policy template argument, \texttt{Kokkos::LaunchBounds<maxT,minB>}. The first argument declares the largest team the kernel will be launched with, the second asks the compiler for a minimum degree of concurrency per compute unit, and thereby caps the register budget. That second argument is backend-specific: blocks per multiprocessor on CUDA, waves per execution unit on HIP. On HIP the compiler must fit each wave into
\begin{equation}
  R_{\mathrm{wave}} \;\le\; \frac{R_{\mathrm{SIMD}}}{\mathtt{minB}} ,
  \label{eq:regbudget}
\end{equation}
with $R_{\mathrm{SIMD}}=512$ vector registers per SIMD on MI250X, and spills whatever does not fit into per-lane private memory. This is distinct from the team scratch of \Cref{sec:opts}: team scratch is the on-chip shared memory the kernel stages its reused data in, whereas the private memory holding spills is backed by device memory. Spilled values are cached on their way there, so a spill is not automatically an HBM access, but it leaves the register file and competes for the same cache capacity as the staged data.

\begin{figure}[t]
  \centering
  \includegraphics[width=\columnwidth]{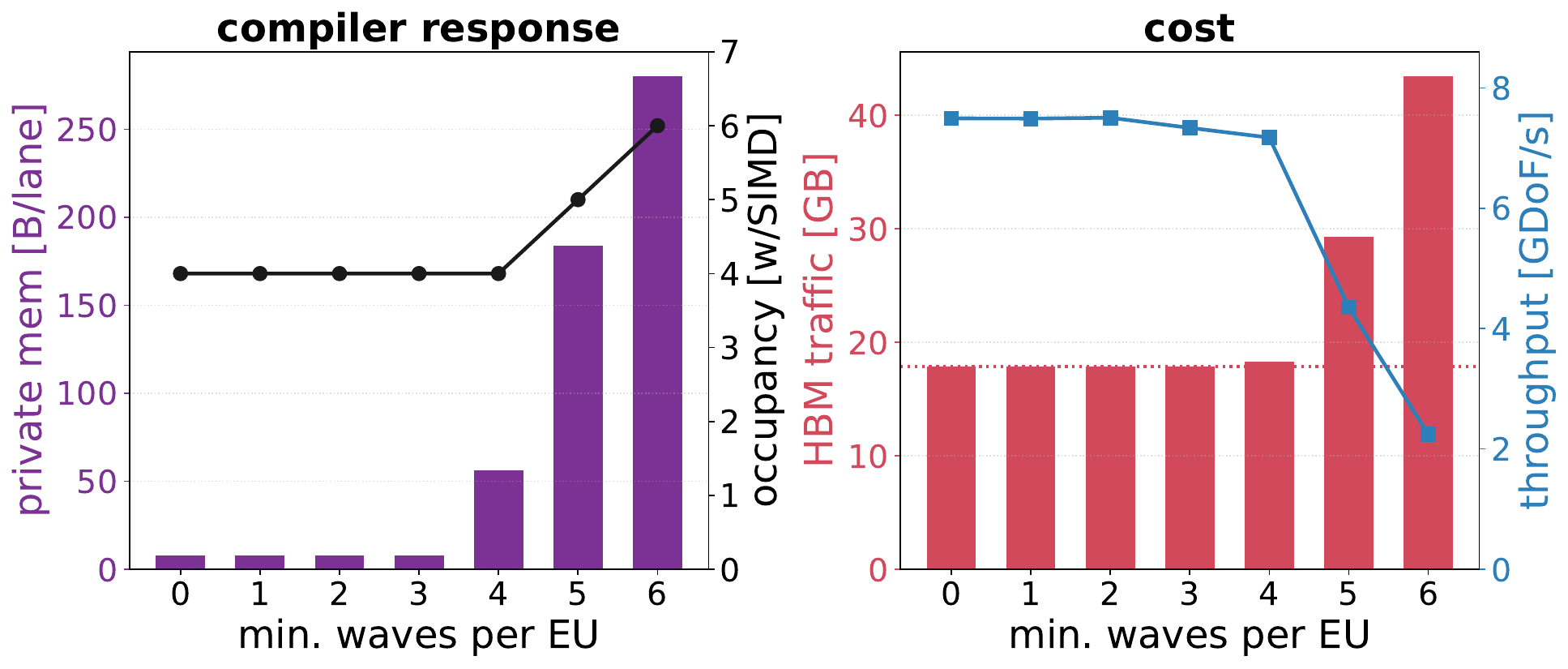}
  \caption{Effect of the launch-bound hint on MI250X at $\ell8$, tile $4/32/1$. Left: per-lane private memory allocated by the compiler and the achieved occupancy in waves per execution unit. The kernel's own private footprint is $8$\,B/lane, everything above that is spilled registers. Right: HBM traffic measured with \texttt{rocprofv3} and throughput.}
  \label{fig:launch-bounds}
\end{figure}

\Cref{fig:launch-bounds} sweeps \texttt{minB} on MI250X. For $\mathtt{minB}\le3$ the bound is not binding: the operator settles at $128$ registers per wave, which by \eqref{eq:regbudget} already admits $512/128=4$ waves, so the setting \texttt{LaunchBounds<512,2>} asks for less than the compiler achieves and is without effect. At $\mathtt{minB}=4$ the cap binds at exactly $128$ registers and spilling to memory begins, but at $56$\,B per lane it stays cheap: HBM traffic rises and throughput falls only slightly. At $5$ and $6$ waves per SIMD, registers per thread tighten to $102$ and $85$, and the compiler meets it by spilling $184$ and $280$ bytes per lane. The kernel then has $64\,\%$ and $143\,\%$ increased HBM traffic. Throughput falls to $58\,\%$ and $30\,\%$ of the spill-free value. The operator is limited by cache traffic rather than by latency, so the extra waves through launch bounds buy nothing while the increased spill traffic worsens the bottleneck.

\subsection{Scaling}
\label{sec:scaling}

\begin{figure*}[t]
  \centering
  \includegraphics[width=\textwidth]{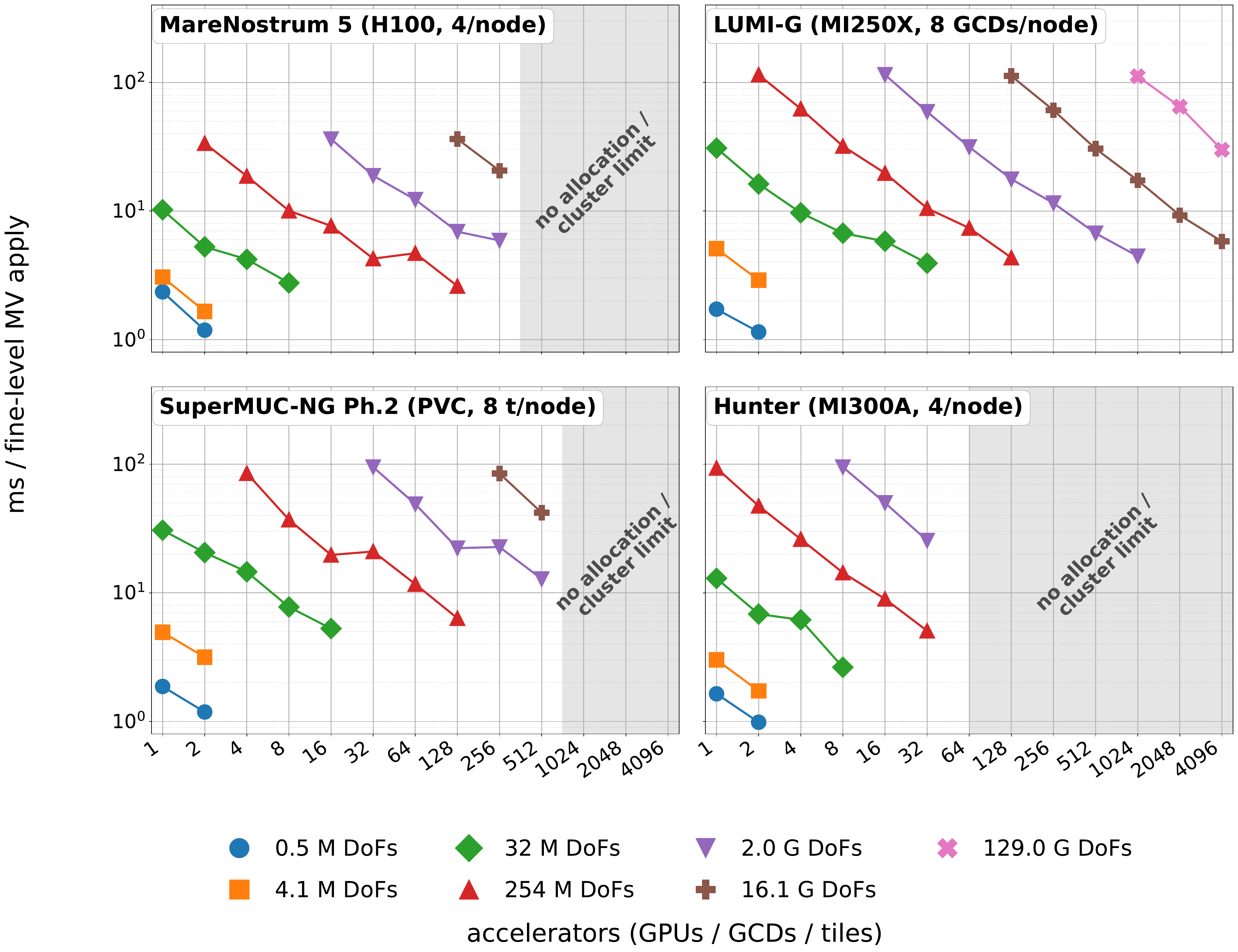}
  \caption{Strong scaling of a single fine-level matrix-free MV apply (the Stokes matrix-vector product inside the Krylov solve, including its halo exchange) of the \texttt{tuned}\,\protect\markStar{} kernel, one line per problem size, labelled by its total number of velocity DoF. The shaded band marks device counts beyond the allocation available on that machine.}
  \label{fig:cross-eps}
\end{figure*}

We run the code on the supercomputers hosting the GPUs considered here: MareNostrum~5 (H100), LUMI-G (MI250X), Hunter (MI300A) and SuperMUC-NG Phase~2 (PVC). \Cref{fig:cross-eps} strong scales the viscous operator on these four machines. Each line holds one problem size fixed and spreads it over more devices, up to the strong-scaling limit at which the halo exchange and the shrinking per-device work stop paying for the added parallelism. On LUMI-G the sweep reaches $4096$ GCDs.

The same plot also carries a weak-scaling reading. Each step to the next model size multiplies the number of DoFs by eight, e.g. from 2 G to 16.1 G DoFs, so pairing it with an eight-fold increase in devices keeps the work per device constant. Read along that horizontal across models, we observe a roughly straight weak-scaling line, most clearly on LUMI-G from 254 M DoFs on $2$ GCDs to 129 G DoFs on $1024$ GCDs. 

\Cref{fig:cross-eps} shows that the kernel is portable and scalable across hundreds of GPUs on all machines.
Scaling of the full coupled simulation, rather than of this single kernel, is reported in the companion paper~\citep{boehm2026terrang}.
\section*{Conclusion}
We compared the performance portability of a low-order, matrix-free finite
element operator across NVIDIA H100, AMD MI250X
and MI300A, and Intel PVC Max~1550 GPUs. The kernel uses team-based GPU dispatch on tiles of elements of the grid, where each team stores shared degrees of freedom from the conforming discretization in scratch memory. We found that also a complex kernel like this can be written once in the hardware-agnostic constructs of Kokkos
and still reach comparable node-level performance on every device, close to the
knee of the roofline and to the attainable bandwidth on most of them. 
One caveat to this is are the few quantities that map directly onto hardware: the
team-tile shape, and the scratch footprint it implies, influences throughput on a range of up to $5.5\times$, and its optimum has to be found manually. However, the optimal tile-shape was more or less universal over the considered cards.

The exception is MI300A, which delivered less throughput than its specifications suggest for our code. As an APU it has to keep atomics coherent with host accesses, so the scatter is resolved in the data fabric rather than in a XCD-local L2, at higher latency.
Accumulating the destination locally before scattering, thereby reducing atomics, recovers $27\,\%$ on MI300A, but costs
$13\,\%$ on MI250X, whose L2 cache already absorbs the atomics. This is a genuine portability
gap: the same code change has opposite impact on two GPUs of the same vendor, and atomics have to be balanced against local accumulation of data and connected spill traffic on the architecture at hand.

\begin{acks}
The authors thank Gabriel Robl and Marcus Mohr for reading and reviewing the
manuscript.

The authors gratefully acknowledge the scientific support and HPC resources
provided by the Erlangen National High Performance Computing Center (NHR@FAU)
of the Friedrich--Alexander--Universit\"at Erlangen--N\"urnberg (FAU). The
hardware is partially funded by the German Research Foundation (DFG).
We acknowledge the EuroHPC Joint Undertaking for awarding this project access to
the EuroHPC supercomputer LUMI, hosted by CSC (Finland) and the LUMI consortium
through a EuroHPC Regular Access call (EHPC-BEN-2026B05-001).

We acknowledge EuroHPC Joint Undertaking for awarding us access to
MareNostrum5 at BSC, Spain (EHPC-BEN-2026B04-050).
The authors gratefully acknowledge the Gauss Centre for Supercomputing e.V.
(\url{www.gauss-centre.eu}) for funding this project by providing computing time
on the GCS Supercomputer SuperMUC-NG at Leibniz Supercomputing Centre
(\url{www.lrz.de}).

The simulations were performed on the national supercomputer HPE Cray EX4000
Hunter at the High Performance Computing Center Stuttgart (HLRS).

\end{acks}

\subsection*{\normalsize\sagesf\bfseries Statements and declarations}

\noindent\textit{Ethical considerations.} This study did not involve human
participants or animals, so ethical approval was not required.

\noindent\textit{Consent to participate.} Not applicable.

\noindent\textit{Consent for publication.} Not applicable.

\begin{dci}
The author(s) declared no potential conflicts of interest with respect to the
research, authorship, and/or publication of this article.
\end{dci}

\begin{funding}
The author(s) disclosed receipt of the following financial support for the
research, authorship, and/or publication of this article: NHR funding is
provided by federal and Bavarian state authorities. NHR@FAU hardware is
partially funded by the German Research Foundation (DFG) -- 440719683.
The authors thank the NHR-Verein e.V.\ for supporting this
work within the NHR Graduate School of National High Performance Computing
(NHR).
\end{funding}

\subsection*{\normalsize\sagesf\bfseries AI disclosure}

\noindent TERRA-NG contains a subset of code that was developed with Claude Code (Anthropic) as an interactive coding assistant, under the close supervision of the authors, who gave detailed instructions and requirements. All functionality, whether written by humans or with machine assistance, has been thoroughly reviewed and tested by the authors, who take full responsibility for it.
During the preparation of this manuscript, the authors used Claude (Anthropic) to draft and improve the readability and language of the text, while all structure and content were provided by the authors. No technical claims, figures or interpretations are AI-generated. The authors reviewed and edited all content and take full responsibility for it.
\subsection*{\normalsize\sagesf\bfseries Data availability statement}

\noindent The source code of \projectname{}, including every operator variant
compared here, is openly available at
\url{https://github.com/mantleconvection/TERRA-NG}.

\bibliographystyle{SageH}
\bibliography{references}

@article{baumgardner1985icos,
  author  = {Baumgardner, J. R. and Frederickson, P. O.},
  title   = {Icosahedral discretization of the two-sphere},
  journal = {SIAM Journal on Numerical Analysis},
  volume  = {22},
  number  = {6},
  pages   = {1107--1115},
  year    = {1985},
  doi     = {10.1137/0722066},
}

@article{zhong2000,
  author  = {Zhong, S. and Zuber, M. T. and Moresi, L. and Gurnis, M.},
  title   = {Role of temperature-dependent viscosity and surface plates in spherical-shell models of mantle convection},
  journal = {Journal of Geophysical Research: Solid Earth},
  volume  = {105},
  number  = {B5},
  pages   = {11063--11082},
  year    = {2000},
  doi     = {10.1029/2000JB900003},
}

@article{zhong2008,
  author  = {Zhong, S. and McNamara, A. and Tan, E. and Moresi, L. and Gurnis, M.},
  title   = {A benchmark study on mantle convection in a 3-{D} spherical shell using {CitcomS}},
  journal = {Geochemistry, Geophysics, Geosystems},
  volume  = {9},
  number  = {10},
  pages   = {Q10017},
  year    = {2008},
  doi     = {10.1029/2008GC002048},
}

@article{tackley2008,
  author  = {Tackley, P. J.},
  title   = {Modelling compressible mantle convection with large viscosity contrasts in a three-dimensional spherical shell using the yin-yang grid},
  journal = {Physics of the Earth and Planetary Interiors},
  volume  = {171},
  number  = {1--4},
  pages   = {7--18},
  year    = {2008},
  doi     = {10.1016/j.pepi.2008.08.005},
}

@article{may2015,
  author  = {May, D. A. and Brown, J. and Le Pourhiet, L.},
  title   = {A scalable, matrix-free multigrid preconditioner for finite element discretizations of heterogeneous {Stokes} flow},
  journal = {Computer Methods in Applied Mechanics and Engineering},
  volume  = {290},
  pages   = {496--523},
  year    = {2015},
  doi     = {10.1016/j.cma.2015.03.014},
}

@article{bauer2019,
  author  = {Bauer, S. and Huber, M. and Ghelichkhan, S. and Mohr, M. and R\"ude, U. and Wohlmuth, B.},
  title   = {Large-scale simulation of mantle convection based on a new matrix-free approach},
  journal = {Journal of Computational Science},
  volume  = {31},
  pages   = {60--76},
  year    = {2019},
  doi     = {10.1016/j.jocs.2018.12.006},
}

@article{kronbichler2019,
  author  = {Kronbichler, M. and Ljungkvist, K.},
  title   = {Multigrid for matrix-free high-order finite element computations on graphics processors},
  journal = {ACM Transactions on Parallel Computing},
  volume  = {6},
  number  = {1},
  pages   = {2:1--2:32},
  year    = {2019},
  doi     = {10.1145/3322813},
}

@article{lin2020viscosity,
  author  = {Lin, Y.-A. and Colli, L. and Wu, J. and Schuberth, B. S. A.},
  title   = {Where are the proto-{South} {China} {Sea} slabs? {SE} {Asian} plate tectonics and mantle flow history from global mantle convection modeling},
  journal = {Journal of Geophysical Research: Solid Earth},
  volume  = {125},
  number  = {12},
  pages   = {e2020JB019758},
  year    = {2020},
  doi     = {10.1029/2020JB019758},
}

@article{stotz2017viscosity,
  author  = {Stotz, I. L. and Iaffaldano, G. and Davies, D. R.},
  title   = {Pressure-driven {Poiseuille} flow: A major component of the torque-balance governing {Pacific} plate motion},
  journal = {Geophysical Research Letters},
  volume  = {45},
  number  = {1},
  pages   = {117--125},
  year    = {2018},
  doi     = {10.1002/2017GL075697},
}

@Article{Kronbichler:2012:CAF,
  author   = {M. Kronbichler and K. Kormann},
  title    = {A generic interface for parallel cell-based finite element operator application},
  journal  = {Computers and Fluids},
  year     = {2012},
  volume   = {63},
  pages    = {135--147},
  doi      = {10.1016/j.compfluid.2012.04.012},
}

@Article{Deubelbeiss:2008:PEPI,
  author       = {Deubelbeiss, Y. and Kaus, B.J.P.},
  title        = {Comparison of Eulerian and Lagrangian numerical techniques for the Stokes equations in the presence of strongly varying viscosity},
  journal      = {Physics of the Earth and Planetary Interiors},
  journaltitle = {Physics of the Earth and Planetary Interiors},
  year         = {2008},
  volume       = {171},
  number       = {1–4},
  month        = dec,
  pages        = {92--111},
  issn         = {0031-9201},
  doi          = {10.1016/j.pepi.2008.06.023},
}

@Article{Heister:2017:GJI,
  author    = {Timo Heister and Juliane Dannberg and Rene Gassmöller and Wolfgang Bangerth},
  title     = {High accuracy mantle convection simulation through modern numerical methods – {II}: realistic models and problems},
  journal   = {Geophysical Journal International},
  year      = {2017},
  volume    = {210},
  number    = {2},
  pages     = {833-851},
  doi       = {10.1093/gji/ggx195},
}

@article{bohm2025code,
  author  = {B\"{o}hm, Fabian and Bauer, Daniel and Kohl, Nils and Alappat, Christie L. and Th\"{o}nnes, Dominik and Mohr, Marcus and K\"{o}stler, Harald and R\"{u}de, Ulrich},
  title   = {Code Generation and Performance Engineering for Matrix-Free Finite Element Methods on Hybrid Tetrahedral Grids},
  journal = {SIAM Journal on Scientific Computing},
  volume  = {47},
  number  = {1},
  pages   = {B131--B159},
  year    = {2025},
  doi     = {10.1137/24M1653756}
}

@book{schubert2001,
  author    = {Schubert, G. and Turcotte, D. L. and Olson, P.},
  title     = {Mantle Convection in the Earth and Planets},
  year      = {2001},
  publisher = {Cambridge University Press},
  address   = {Cambridge},
  doi       = {10.1017/CBO9780511612879}
}

@misc{boehm2026terrang,
  author        = {B\"ohm, F. and Kohl, N. and Ilangovan, P. and Robl, G. and Rezaei, F. and Mohr, M. and Schuberth, B. S. A. and K\"ostler, H. and Bunge, H.-P. and R\"ude, U.},
  title         = {{TERRA-NG} v1.0: Extreme-Scale, {GPU}-accelerated Mantle Convection},
  year          = {2026},
  eprint        = {2609.21633},
  archivePrefix = {arXiv},
  primaryClass  = {cs.CE},
  doi           = {10.48550/arXiv.2609.21633},
  note          = {arXiv preprint arXiv:2609.21633},
}

@article{settgast2023lowordergpu,
  author  = {Settgast, Randolph R. and Dudouit, Yohann and Castelletto, Nicola and Tobin, William R. and Corbett, Benjamin C. and Klevtsov, Sergey},
  title   = {Performant low-order matrix-free finite element kernels on {GPU} architectures},
  journal = {arXiv preprint},
  volume  = {arXiv:2308.09839},
  year    = {2023},
  doi     = {10.48550/arXiv.2308.09839},
}

@article{demeshko2019albany,
  author  = {Demeshko, Irina and Watkins, Jerry and Tezaur, Irina K. and Guba, Oksana and Spotz, William F. and Salinger, Andrew G. and Pawlowski, Roger P. and Heroux, Michael A.},
  title   = {Toward performance portability of the {Albany} finite element analysis code using the {Kokkos} library},
  journal = {The International Journal of High Performance Computing Applications},
  volume  = {33},
  number  = {2},
  pages   = {332--352},
  year    = {2019},
  doi     = {10.1177/1094342017749957},
}

@article{fischer2020scalability,
  author  = {Fischer, Paul and Min, Misun and Rathnayake, Thilina and Dutta, Som and Kolev, Tzanio and Dobrev, Veselin and Camier, Jean-Sylvain and Kronbichler, Martin and Warburton, Tim and {\'S}wirydowicz, Kasia and Brown, Jed},
  title   = {Scalability of high-performance {PDE} solvers},
  journal = {The International Journal of High Performance Computing Applications},
  volume  = {34},
  number  = {5},
  pages   = {562--586},
  year    = {2020},
  doi     = {10.1177/1094342020915762},
}

@article{chalmers2023hipbone,
  author  = {Chalmers, Noel and Mishra, Abhishek and McDougall, Damon and Warburton, Tim},
  title   = {{HipBone}: A performance-portable graphics processing unit-accelerated {C++} version of the {NekBone} benchmark},
  journal = {The International Journal of High Performance Computing Applications},
  volume  = {37},
  number  = {5},
  pages   = {560--577},
  year    = {2023},
  doi     = {10.1177/10943420231178552},
}

@article{vargas2022mfem,
  author  = {Vargas, Arturo and Stitt, Thomas M. and Weiss, Kenneth and Tomov, Vladimir Z. and Camier, Jean-Sylvain and Kolev, Tzanio and Rieben, Robert N.},
  title   = {Matrix-free approaches for {GPU} acceleration of a high-order finite element hydrodynamics application using {MFEM}, {Umpire}, and {RAJA}},
  journal = {The International Journal of High Performance Computing Applications},
  volume  = {36},
  number  = {4},
  pages   = {492--509},
  year    = {2022},
  doi     = {10.1177/10943420221100262},
}

@article{kolev2021efficient,
  author  = {Kolev, Tzanio and Fischer, Paul and Min, Misun and Dongarra, Jack and Brown, Jed and Dobrev, Veselin and Warburton, Tim and Tomov, Stanimire and Shephard, Mark S. and Abdelfattah, Ahmad and others},
  title   = {Efficient exascale discretizations: High-order finite element methods},
  journal = {The International Journal of High Performance Computing Applications},
  volume  = {35},
  number  = {6},
  pages   = {527--552},
  year    = {2021},
  doi     = {10.1177/10943420211020803},
}

@article{bertagna2019hommexx,
  author  = {Bertagna, Luca and Deakin, Michael and Guba, Oksana and Sunderland, Daniel and Bradley, Andrew M. and Tezaur, Irina K. and Taylor, Mark A. and Salinger, Andrew G.},
  title   = {{HOMMEXX} 1.0: a performance-portable atmospheric dynamical core for the {Energy Exascale Earth System Model}},
  journal = {Geoscientific Model Development},
  volume  = {12},
  number  = {4},
  pages   = {1423--1441},
  year    = {2019},
  doi     = {10.5194/gmd-12-1423-2019},
}

@article{arndt2025dealii,
  author  = {Africa, Pasquale C. and Arndt, Daniel and Bangerth, Wolfgang and Blais, Bruno and Fehling, Marc and Gassm{\"o}ller, Rene and Heister, Timo and Heltai, Luca and Kinnewig, Sebastian and Kronbichler, Martin and Maier, Matthias and Munch, Peter and Schreter-Fleischhacker, Magdalena and Thiele, Jan P. and Turcksin, Bruno and Wells, David and Yushutin, Vladimir},
  title   = {The {deal.II} library, Version 9.6},
  journal = {Journal of Numerical Mathematics},
  volume  = {32},
  number  = {4},
  pages   = {369--380},
  year    = {2024},
  doi     = {10.1515/jnma-2024-0137},
}

@article{johansson2025lammps,
  author  = {Johansson, Anders and Weinberg, Evan and Trott, Christian R. and McCarthy, Megan J. and Moore, Stan G.},
  title   = {{LAMMPS-KOKKOS}: Performance Portable Molecular Dynamics Across Exascale Architectures},
  journal = {arXiv preprint arXiv:2508.13523},
  year    = {2025},
  doi     = {10.48550/arXiv.2508.13523},
}

@article{asahi2025kokkosfft,
  author  = {Asahi, Yuuichi and Padioleau, Thomas and Zehner, Paul and Bigot, Julien and Lebrun-Grandi{\'e}, Damien},
  title   = {kokkos-fft: A shared-memory {FFT} for the {Kokkos} ecosystem},
  journal = {Journal of Open Source Software},
  volume  = {10},
  number  = {111},
  pages   = {8391},
  year    = {2025},
  doi     = {10.21105/joss.08391},
}

@article{bauer2025multigrid,
  author  = {Bauer, Daniel and Kohl, Nils and McCormick, Stephen F. and Tamstorf, Rasmus},
  title   = {Multigrid with Linear Storage Complexity},
  journal = {arXiv preprint arXiv:2511.19036},
  year    = {2025},
  url     = {https://arxiv.org/abs/2511.19036},
}

@article{tamstorf2021discretization,
  author  = {Tamstorf, Rasmus and Benzaken, Joseph and McCormick, Stephen F.},
  title   = {Discretization-Error-Accurate Mixed-Precision Multigrid Solvers},
  journal = {SIAM Journal on Scientific Computing},
  volume  = {43},
  number  = {5},
  pages   = {S420--S447},
  year    = {2021},
  doi     = {10.1137/20M1349230},
}

@article{trott2022kokkos,
  author  = {Trott, Christian R. and Lebrun-Grandi{\'e}, Damien and Arndt, Daniel and Ciesko, Jan and Dang, Vinh and Ellingwood, Nathan and Gayatri, Rahulkumar and Harvey, Evan and Hollman, Daisy S. and Ibanez, Dan and Liber, Nevin and Madsen, Jonathan and Miles, Jeff and Poliakoff, David and Powell, Amy and Rajamanickam, Sivasankaran and Simberg, Mikael and Sunderland, Dan and Turcksin, Bruno and Wilke, Jeremiah},
  title   = {{Kokkos} 3: Programming Model Extensions for the Exascale Era},
  journal = {IEEE Transactions on Parallel and Distributed Systems},
  volume  = {33},
  number  = {4},
  pages   = {805--817},
  year    = {2022},
  doi     = {10.1109/TPDS.2021.3097283},
}

@article{edwards2014kokkos,
  author  = {Edwards, H. Carter and Trott, Christian R. and Sunderland, Daniel},
  title   = {{Kokkos}: Enabling manycore performance portability through polymorphic memory access patterns},
  journal = {Journal of Parallel and Distributed Computing},
  volume  = {74},
  number  = {12},
  pages   = {3202--3216},
  year    = {2014},
  doi     = {10.1016/j.jpdc.2014.07.003},
}

\end{document}